\documentclass[
 amsmath,amssymb,
 aps,
 twocolumn,
 pra,
 superscriptaddress
]{revtex4-2}
\usepackage{amsthm}
\usepackage{manu}
\usepackage{graphicx} 
\usepackage{dcolumn}
\usepackage{bm}
\usepackage{mathdots}
\usepackage[colorlinks=true,linkcolor=red,citecolor=blue,urlcolor=blue,]{hyperref}
\usepackage{xcolor, color}
\usepackage{xspace}
\usepackage{extarrows}
\usepackage{makecell}
\usepackage{physics}
\usepackage{ulem}

\begin{document}

\title{Inverse designed Hamiltonians for perfect state transfer and remote entanglement generation, and applications in superconducting qubits}
\author{Tian-Le Wang}
\addrLab\addrSyn
\author{Ze-An Zhao}
\addrLab\addrSyn
\author{Peng Wang}
\addrLab\addrSyn\addrSuzhou
\author{Sheng Zhang}
\addrLab\addrSyn\addrSuzhou
\author{Ren-Ze Zhao}
\author{Xiao-Yan Yang}
\author{Hai-Feng Zhang}
\author{Zhi-Fei Li}
\author{Yuan Wu}
\addrLab\addrSyn

\author{Peng Duan}
\email{pengduan@ustc.edu.cn}
\addrLab\addrSyn

\author{Ming Gong}
\email{gongm@ustc.edu.cn}
\addrLab\addrSyn
\addrHefei
\addrAnhui

\author{Guo-Ping Guo}
\email{gpguo@ustc.edu.cn}
\addrLab\addrSyn\addrOrigin

\date{\today}
\begin{abstract}
    Hamiltonian inverse engineering enables the design of protocols for specific quantum evolutions or target state preparation. Perfect state transfer (PST) and remote entanglement generation are notable examples, as they serve as key primitives in quantum information processing. 
    However, Hamiltonians obtained through conventional methods often lack robustness against noise. 
    Assisted by inverse engineering, we begin with a noise-resilient energy spectrum and construct a class of Hamiltonians, referred to as the dome model, that significantly improves the system's robustness against noise, as confirmed by numerical simulations. 
    This model introduces a tunable parameter $m$ that modifies the energy-level spacing and gives rise to a well-structured Hamiltonian. It reduces to the conventional PST model at $m=0$ and simplifies to a SWAP model involving only two end qubits in the large-$m$ regime. 
    To address the challenge of scalability, we propose a cascaded strategy that divides long-distance PST into multiple consecutive PST steps. Our work is particularly suited for demonstration on superconducting qubits with tunable couplers, which enable rapid and flexible Hamiltonian engineering, thereby advancing the experimental potential of robust and scalable quantum information processing.  
\end{abstract}

\maketitle

\section{Introduction}
    Perfect state transfer (PST)~\cite{2003_PST_Bose, 2004_PST_Christandl, 2010_PST_Kay, 2005_PST_YungManHong, 2012_PST_Vinet_PRA, 2012_pst_vinet_jpa} enables the establishment of transmission channels in quantum devices with only nearest-neighbor interactions. By pre-engineering the global Hamiltonian, a quantum state can be perfectly transferred from one end of the channel to the other across multiple qubits, thereby enhancing device connectivity. Fractional state transfer (FST)~\cite{2016_FST_Genest, 2016_FST_Lemay, 2017_Lanczos_quantum_Kay, 2017_Lanczos_njp_Kay}, an extension of PST, permits fractional revival of the quantum state at both ends of the channel after evolution, with controllable amplitudes and relative phase, making remote entanglement generation feasible. As fundamental building blocks for on-chip quantum interconnects and modular quantum computing~\cite{2024_Wallraff_self-testing, 2023_Wallraff_Loophole-free, 2020_Wallraff_Quantum-Link, 2024_remote-gates_YuHaifeng, 2024_RIP_YuYang, 2024_plug-play_Michael, 2024_ZhongYoupeng_Anyonic, 2023_ZhongYoupeng_teleportation, 2023_ZhongYoupeng_Low-loss, 2021_ZhongYoupeng_deterministic, 2021_entangle-die_Gold, 2022_chiplet_Smith}, both PST and remote entanglement generation are becoming increasingly important as quantum chip scale continues to grow~\cite{2024_Malekakhlagh_EnhancedPST, 2025_newest_willow, 2019_newest_Sycamore, 2025_newest_Zuchongzhi3.1, 2025_newest_Zuchongzhi3.0, 2022_newest_Zuchongzhi2.1, 2021_newest_Zuchongzhi2.0, 2021_newest_Zuchongzhi1.0, 2025_newest_ZJU_11x11, 2025_newest_ZJU_2x20, 2024_newest_ZJU, 2022_newest_ZJU, 2025_newest_Chuang-tzu2.0, 2023_newest_Chuang-tzu, 2025_newest_FanHeng, 2024_newest_IBM, 2023_newest_IBM, 2022_newest_IBM, 2024_newest_baqis, 2025_newest_YuDapeng, 2023_newest_YuDapeng, 2021_newest_jiuzhang2.0, 2020_newest_jiuzhang1.0, 2023_newest_ions, 2024_newest_atoms}.

    In both PST and FST, the focus is typically on the qubits directly involved in the task. However, in solid-state quantum devices with static qubits~\cite{2025_newest_willow, 2025_newest_Zuchongzhi3.0, 2024_newest_IBM, 2021_NV_Pezzagna, semiconductor_burkard_2023, semiconductor_stano_2022}, quantum information processing between targeted qubits inevitably relies on intermediate qubits as transmission media. In most experimental implementations~\cite{2018_PST_LiX, 2023_PST_ZhangChi, 2024_PST_XiangLiang, 2024_ParityPST_Roy, 2016_optical_Chapman, 2013_optical_perez, 2012_optical_bellec, 2005_NMR_ZhangJingfu, 2020_nanoelectromechanical_TianTian}, these intermediate qubits remain populated throughout the process, introducing multiple noise channels that substantially degrade fidelity. Although some adiabatic protocols can suppress population on the intermediate qubits by constructing dark states~\cite{2020_adiabatic_ChangHS, 2017_stirap_Vitanov, 2013_stirap_Sangchul, 2017_stirap_WangYingDan}, they demand highly precise temporal control of pulses. Consequently, the design of pre-engineered Hamiltonians~\cite{inverse_2017_wang, inverse_2018_chertkov, inverse_2023_inui, inverse_2024_inui, inverse_ashhab_2015, inverse_ji_2024, inverse_nelmes_2025, inverse_faria_2025} with intrinsic noise resilience has emerged as an important research direction. 

    The Hamiltonians for PST and FST typically exhibit specific structural properties~\cite{2016_FST_Genest, 2012_PST_Vinet_PRA, 2010_PST_Kay}. First, they must be tridiagonal, as required by the qubit chain architecture, where only nearest-neighbor couplings are permitted. Second, the observation that quantum states in PST (or FST) exclusively transfer (or entangle) between symmetric sites implies that the Hamiltonian possesses specific symmetries or eigenvalue relationships. 
    Conversely, if one could reconstruct PST and FST Hamiltonians from desired dynamical properties or eigenvalue relations, this would provide a more application-driven perspective and potentially reveal new Hamiltonian models. 
    It is worth pointing out that in matrix analysis there exists a well-established class of inverse problems for Jacobi (tridiagonal) matrices, where the matrix can be reconstructed from a prescribed eigenvalue spectrum~\cite{2006_iep_gladwell, 2005_iep_Chu}. When the Jacobi matrix is constrained to satisfy mirror symmetry, the reconstructed form is unique. This structural property motivates us to apply these results to the reconstruction of PST and FST Hamiltonians. 
    Here, mirror symmetry means that the qubit frequencies and coupling strengths are symmetric with respect to the center of the transmission chain, i.e., $\omega_n = \omega_{N+1-n}$ for $n=1,2,\cdots,N$, and $\ J_n = J_{N-n}$ for $n=1,2,\cdots,N-1$ in a $1 \times N$ qubit chain. 
    Such a configuration is readily implementable on solid-state quantum platforms such as superconducting quantum circuits~\cite{2023_coupler_ZhangChi, 2019_coupler_krantz, 2018_coupler_yanfei}.

    \begin{table*} 
    \caption{
    Summary of experimental demonstrations of PST and/or FST across different research groups and hardware platforms. 
    }
    \renewcommand{\arraystretch}{1.5}
    \begin{tabular}{p{0.25\textwidth}|p{0.15\textwidth}|p{0.12\textwidth}|p{0.45\textwidth}}
    \hline
    \textbf{Reference} & \textbf{Physical system} & \textbf{System size} & \textbf{Results (Fidelity, quantum state and  time scale)} \\
    \hline
    Wang \textit{et al.} (2026)~\cite{gpst_wang_2025} & Transmon qubits & \makecell[l]{$1\times5$ \\ $3\times3$} & \makecell[l]{92.5\% for Bell state generation (57~ns)~\footnotemark[1] \\ 85.0\% for $\mathcal{W}$ state generation (56~ns)~\footnotemark[1]} \\
    \hline
    Roy \textit{et al.} (2025)~\cite{2024_ParityPST_Roy} & Transmon qubits & 1 × 6 & 88.1\% for GHZ state generation (390~ns)~\footnotemark[1] \\
    \hline
    Xiang \textit{et al.} (2024)~\cite{2024_PST_XiangLiang} & Transmon qubits & 6 × 6 & \makecell[l]{90.2\% for single-excitation transfer (250~ns)~\footnotemark[2] \\ 84.0\% for Bell state transfer (250~ns)~\footnotemark[1] \\ 73.7\% for two-excitation transfer (250~ns)~\footnotemark[2]} \\
    \hline
    Zhang \textit{et al.} (2023)~\cite{2023_PST_ZhangChi} & Transmon qubits & 1 × 4 & 98.6\% for single-excitation transfer (25~ns)~\footnotemark[4] \\
    \hline
    Li \textit{et al.} (2018)~\cite{2018_PST_LiX} & Transmon qubits & 1 × 4 & 99.2\% for single-excitation transfer (84~ns)~\footnotemark[4] \\
    \hline
    Chapman \textit{et al.} (2016)~\cite{2016_optical_Chapman} & Optical waveguide & 1 × 11 & \makecell[l]{98.2\% for single-excitation transfer~\footnotemark[3] \\ 97.1\% for Bell state transfer~\footnotemark[5]}\\
    \hline
    \end{tabular}

    \footnotetext[1]{
    Fidelity is defined by quantum state tomography (QST)~\cite{qst_2012_smolin}, $F_{\text{state}}=\mathrm{Tr}(\rho_{\text{exp}} \rho_{\text{ideal}})$. $\rho_{\text{exp}}$ is the density matrix reconstructed from the QST experiment, and $\rho_{\text{ideal}}$ is the ideal density matrix. 
    }
    \footnotetext[2]{
    Transfer fidelity, the population probability of the final states~\cite{2024_PST_XiangLiang}.
    }
    \footnotetext[3]{
    Fidelity obtained via quantum process tomography (QPT)~\cite{qpt_2018_knee, qpt_2013_merkel, qpt_2013_korotkov}, $F_{\text{process}}=\mathrm{Tr}(\chi_{\text{exp}} \chi_{\text{ideal}})$. $\chi_{\text{exp}}$ is the process matrix reconstructed from the QPT experiment, and $\chi_{\text{ideal}}$ is the ideal quantum process matrix. 
    }
    \footnotetext[4]{
    Fidelity is extracted from curve fitting of parameter $P$ using the model $F=A P^m + B$, where $A, B$ and $P$ are fitting parameters, and $F$ is the quantum process fidelity after $m$ rounds of transfer.
    }
    \footnotetext[5]{
    Fidelity calculated from two-qubit polarization tomography~\cite{polartomo_james_2001}, $F_{2Q}=\mathrm{Tr}(\sqrt{\sqrt{\rho_{\text{exp}}} \rho_{\text{ref}} \sqrt{\rho_{\text{exp}}}})^2$. $\rho_{\text{exp}}$ is the density matrix after PST process, $\rho_{\text{ref}}$ is the density matrix after a reference experiment. 
    } \label{tab-review}
    \end{table*}

    In this work, we analytically derive a class of noise-resilient Hamiltonians, referred to as the dome model, using the inverse eigenvalue method (IEM) for both FST and PST. Specifically, by introducing a tunable parameter $m$ to control the energy gap, we construct a novel eigenvalue spectrum. As $m$ increases, the population of intermediate qubits is significantly suppressed during evolution, leading to improved noise robustness. This phenomenon can be understood from the perspective of the effective Hamiltonian in the large-$m$ limit. Furthermore, the dome model provides a unified framework that enables the sequential realization of FST and PST within a single evolution period. 

    The remainder of the paper is organized as follows. Sec.~\ref{sec-review} revisits PST and FST from an eigenvalue perspective, leading to a spectral viewpoint that ultimately motivates our model. Sec.~\ref{sec-dome} constructs the dome Hamiltonian using the inverse eigenvalue method summarized in Appendix~\ref{app:iem}. We then systematically investigate its dynamics (Sec.~\ref{sec-dynamic}), robustness against both coherent and decoherence noise (Sec.~\ref{sec-noise}), and scalability (Sec.~\ref{sec-scale}). Sec.~\ref{sec-longpst} presents a proposal for realizing long-distance PST. Finally, Sec.~\ref{sec-conclusion} summarizes the paper and outlines potential directions for future research. 

   \begin{figure}
       \centering
       \includegraphics[width=0.75\columnwidth]{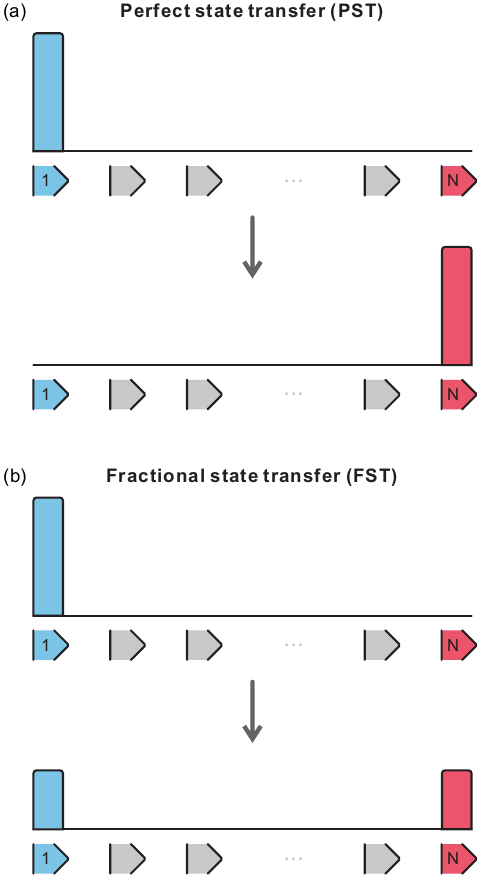}
       \caption{
       Schematic of perfect state transfer (PST) and fractional state transfer (FST). 
       (a) In PST, a quantum state initialized at $\ket{1}$ (or more generally, $\ket{n}$) is perfectly transferred to $\ket{N}$ ($\ket{N-n+1}$) after a transfer time $\tau_{\text{PST}}$. 
       (b) In FST, a quantum state initialized at $\ket{1}$ ($\ket{n}$) evolves into a superposition of $\ket{1}$ and $\ket{N}$ ($\ket{n}$ and $\ket{N-n+1}$) after a certain time $\tau_{\text{FST}}$. The amplitudes and relative phase between $\ket{1}$ and $\ket{N}$ ($\ket{n}$ and $\ket{N-n+1}$) can be controlled by engineering the system Hamiltonian. 
       }
       \label{fig-fig1}
   \end{figure}

\section{Revisiting PST and FST from an Eigenvalue Perspective }\label{sec-review}
    In this section, after a brief introduction of PST and FST, we revisit these processes from an eigenvalue perspective. This viewpoint highlights the spectral structures underlying state-transfer dynamics and naturally motivates the construction of the new model introduced in Sec.~\ref{sec-dome}. We begin with the PST scheme proposed in Ref.~\onlinecite{2004_PST_Christandl}, where the Hamiltonian of a  1D chain of $N$ qubits is given by
    \begin{equation}
    H = \sum_{n=1}^{N-1} \frac{J_n}{2} \left(\sigma_n^x \sigma_{n+1}^x + \sigma_n^y \sigma_{n+1}^y \right),
    \end{equation}
    where $\sigma_n^{x,y}$ are the Pauli operators and the coupling strengths $J_n$ satisfy  
    \begin{equation}
    J_n = \frac{J}{2}\sqrt{n(N-n)}, \label{eq-Jn}
    \end{equation}
    where $J$ denotes the evolution rate of the system, so that the evolution period is $T = 2\pi / J$. Since all qubit frequencies are resonant, we refer to this type of configuration as the line model. 

    When considering only one excitation, the dynamics is restricted to the subspace $S_G$, where in a chain of length $N$, only one spin points up and the remaining $N-1$ spins point down. In this regime, the spin operators map to fermionic operators as $\sigma_n^+ \sim c_n^\dagger$ and $\sigma_n^- \sim c_n$, yielding
    \begin{equation}
    H = \sum_{n=1}^{N-1} J_n (c_n^\dagger c_{n+1} + \text{h.c.}).
    \end{equation}
    Here, h.c. denotes the Hermitian conjugate.
    There is an intimate relation between $J_n$ and the matrix elements of the spin-$x$ operator, which satisfy 
    \begin{align}
        S_x = \frac{1}{2}\sum_{m_s=-S}^{S-1} &\sqrt{S(S+1) - m_s(m_s+1)}  \nonumber\\
        &\times\left(\ket{m_s+1}\bra{m_s} + \text{h.c.}\right),
    \end{align}
    these two expressions become equivalent under the substitution $n = S + m_s + 1$, and $N = 2S+1$. In this way, if we represent the spin using the basis $\ket{\downarrow \downarrow \cdots \downarrow \uparrow \downarrow \cdots \downarrow}$, we will find that the Hamiltonian is given by 
    \begin{equation}
    H = JS_x,
    \label{eq-Sx}
    \end{equation}
    whose eigenvalues are equally spaced and given by $\{-S, -S+1, -S+2, \cdots, S\}\times J$. 
    This uniform spectrum is the key principle that enables PST in Ref.~\onlinecite{2004_PST_Christandl}. This idea can be extended to other many-body models
    \begin{equation}
    H = \sum_{n=1}^{N-1} \frac{J_n}{2} \left(\sigma_n^x \sigma_{n+1}^x + \sigma_n^y \sigma_{n+1}^y \right) + \sum_{n=1}^{N} \frac{\omega_n}{2} \sigma_n^z.
    \label{eq-H}
    \end{equation}
    With appropriately chosen $\omega_n$ and $J_n$, the model reproduces the same state transfer behavior as that of the spin model discussed above~\cite{2004_PST_Christandl, 2010_PST_Kay, 2012_PST_Vinet_PRA, 2012_pst_vinet_jpa, 2016_FST_Genest}. 

    Before imposing specific state transfer conditions, we first establish the spectral structure implied by the nearest-neighbor Hamiltonian in Eq.~\eqref{eq-H}. 
    Since $H$ conserves the total excitation number, its dynamics can be restricted to the single-excitation subspace spanned by the site states $\ket{n} = \sigma_n^+ \ket{\downarrow \downarrow \downarrow\cdots  \downarrow}\, (n=1,2,3,\cdots,N)$. 
    Let $\{\ket{\lambda_s}\}_{s=1}^{N}$ denote the orthonormal eigenbasis of $H$, satisfying $H\ket{\lambda_s}=\lambda_s\ket{\lambda_s}$, where the corresponding eigenvalues are arranged in ascending order as $\lambda_1<\lambda_2<\cdots<\lambda_N$. 
    Each site state $\ket{n}$ can be expressed in this eigenbasis as
    \begin{equation}
        \ket{n} = \sum_{s=1}^N W_{sn} \ket{\lambda_s}, 
        \label{eq-ketn}
    \end{equation}
    where $W_{sn}=\langle\lambda_s|n\rangle$ denotes the transformation coefficient between the single-excitation state $\ket{n}$ and the eigenbasis $\ket{\lambda_s}$. 
    As a consequence of the tridiagonal structure of the Hamiltonian in Eq.~\eqref{eq-H}, $W_{sn}$ can be factorized into the product of an $s$-dependent weight term and an $n$-dependent polynomial evaluated at $\lambda_s$. To establish this result explicitly, we note that in the single-excitation subspace,
    \begin{equation}
        H\ket{n}
        =J_{n-1}\ket{n-1}+\omega_n\ket{n}+J_n\ket{n+1},
    \end{equation} 
    the above relation can be rearranged as
    \begin{equation}
        J_n\ket{n+1}=(H-\omega_n)\ket{n}-J_{n-1}\ket{n-1}.
    \end{equation}
    For a connected chain, $J_n \neq 0$ for $1 \leq n \leq N-1$. Starting from $\ket{1}$ and applying this relation iteratively, we derive
    \begin{align}
        \ket{2} &= \frac{H - \omega_1}{J_1} \ket{1}, \\
        \ket{3} &= \frac{(H - \omega_2)(H - \omega_1) - J_1^2}{J_1 J_2} \ket{1}, \\
        \ket{n} &= \chi_n(H)\ket{1}. \label{eq-ketn-chin-ket1}
    \end{align}
    Here, $\chi_n(H)$ is a polynomial of degree $(n-1)$ in $H$, with the initial condition $\chi_1(H)=1$. Using the spectral decomposition $H = \sum_{s=1}^N \lambda_s \ket{\lambda_s}\bra{\lambda_s}$, $\chi_n(H)$ can be written as
    \begin{equation}
        \chi_n(H) = \sum_{s=1}^N \chi_n(\lambda_s) \ket{\lambda_s}\bra{\lambda_s}.
        \label{eq-chin-expan}
    \end{equation}
    $\chi_n(\lambda_s)$ is a real polynomial of degree $(n-1)$ with $\chi_1(\lambda_s) = 1$. A more detailed analysis of these polynomials is provided in Appendix~\ref{app:iem}. Substituting Eq.~\eqref{eq-chin-expan} into Eq.~\eqref{eq-ketn-chin-ket1}, 
    and expanding $\ket{1}$ in the eigenbasis as $\ket{1} = \sum_{s=1}^N W_{s1} \ket{\lambda_s}$, 
    we find
    \begin{align}
        \ket{n} 
        = \sum_{s=1}^N \chi_n(\lambda_s) \ket{\lambda_s} \braket{\lambda_s}{1} = \sum_{s=1}^N W_{s1} \chi_n(\lambda_s) \ket{\lambda_s},
    \end{align}
    we fix the overall phase of each eigenvector such that its component at the first site is real and positive, i.e., $W_{s1} = \braket{\lambda_s}{1} = \braket{1}{\lambda_s} > 0$ for $s=1,2,\dots, N$. Defining the corresponding endpoint weight as $\alpha_s \equiv \abs{W_{s1}}^2 > 0$, it then follows that $W_{s1} = \sqrt{\alpha_s}$ and hence
    \begin{equation}
        \ket{n} = \sum_{s=1}^N \sqrt{\alpha_{s}} \chi_n(\lambda_s) \ket{\lambda_s}.
        \label{eq-ketn-chin}
    \end{equation}
    Comparing Eq.~\eqref{eq-ketn-chin} with Eq.~\eqref{eq-ketn} yields $W_{sn}=\sqrt{\alpha_s}\,\chi_n(\lambda_s)$. Thus, each transformation coefficient is expressed as a product of the square root of the weight $\sqrt{\alpha_s}$ and the polynomial $\chi_n(\lambda_s)$ evaluated at $\lambda_s$.
    From $\braket{m}{n} = \delta_{mn}$, we obtain $\sum_s \alpha_s \chi_m(\lambda_s)\chi_n(\lambda_s) = \delta_{mn}$. Polynomials satisfying this orthogonality relation are referred to as orthogonal polynomials~\cite{2016_FST_Genest, 2006_iep_gladwell}. Their properties have been extensively studied and will serve as a central theoretical tool for Hamiltonian reconstruction, which will be discussed in detail in Appendix~\ref{app:iem}. 

    We now analyze the FST process from the wavefunction perspective, as schematically illustrated in Fig.~\ref{fig-fig1}. Building on this framework, we first derive the necessary and sufficient conditions for PST, and then discuss two special cases of the FST process. Specifically, we consider the evolution of an initial state $\ket{1}$ into a superposition of $\ket{1}$ and $\ket{N}$ after a time interval $\tau$, expressed as 
    \begin{align}
        e^{-iH \tau} \ket{1} &= e^{i\phi} (\sin{\theta}\ket{1} + e^{i\psi} \cos{\theta}\ket{N}). \label{eq-FST}
    \end{align}
    
    Expanding Eq.~\eqref{eq-FST} using the eigenbasis from Eq.~\eqref{eq-ketn-chin} yields 
    \begin{align}
        e^{-i\lambda_s \tau} &= e^{i\phi} (\sin{\theta} + e^{i\psi} \cos{\theta}\chi_N(\lambda_s)), \label{eq-expansion}
    \end{align}
    where we use the initial condition $\chi_1(\lambda_s)=1$. When $\theta=0$, Eq.~\eqref{eq-expansion} reduces to the PST case, i.e., 
    \begin{align}
        e^{-i\lambda_s \tau} &= e^{i\phi} \chi_N(\lambda_s), \label{eq-theta0}
    \end{align}
    where the phase factor $e^{i\psi}$ in Eq.~\eqref{eq-expansion} has been absorbed into the global phase.

    According to the theory of orthogonal polynomials, $\chi_n(\lambda)$ is real-valued~\cite{2016_FST_Genest, 2006_iep_gladwell}, which together with Eq.~\eqref{eq-theta0}, implies that $\chi_N(\lambda_s) = \pm 1$. Moreover, the zeros of $\chi_N(\lambda)$ must lie between those of $\chi_{N+1}(\lambda)$. As shown in Appendix~\ref{app:iem}, $\chi_{N+1}(\lambda) \propto  \prod_{s=1}^N (\lambda-\lambda_s)$, which restricts $\chi_N(\lambda_s)$ to the forms $(-1)^{N+s}$ or $(-1)^{N+s+1}$. Combining this with the positivity condition $\alpha_s > 0$ in Eqs.~\eqref{eq-recurr_weight} and \eqref{eq-vec_prin}, we obtain
    \begin{equation}
        \chi_N(\lambda_s) = (-1)^{N+s}. \label{eq-chiN-sign}
    \end{equation} 
    Using Eq.~\eqref{eq-chiN-sign} together with Eq.~\eqref{eq-ketn-chin}, one obtains
    \begin{equation}
        \ev{H^x}{1} = \sum_{s=1}^N \alpha_s \lambda_s^x = \ev{H^x}{N}, \label{eq-mirror_H}
    \end{equation}
    where $H^x$ denotes the $x$-th power of $H$, and $x$ is a positive integer. For $x=1$, Eq.~\eqref{eq-mirror_H} directly leads to $\omega_1 = \omega_N$. For $x=2$, $J_1$ and $J_{N-1}$ appear on the two sides of the equation, giving $\omega_1^2 + J_1^2 = \omega_N^2 + J_{N-1}^2$. Thus, $J_1^2 = J_{N-1}^2$. Repeating this procedure for higher powers of $H$ successively introduces the remaining Hamiltonian parameters and ultimately yields the mirror symmetry conditions~\cite{2010_PST_Kay}
    \begin{subequations}
    \begin{align}
        \omega_n &= \omega_{N+1-n}, \qquad n=1,2,\cdots, N,\\
        \quad J_n &= J_{N-n}, \qquad n=1,2,\cdots, N-1.
    \end{align}
    \label{eq-mirror_condition}
    \end{subequations}
    In deriving the above relations, we have used the fact that all $J_n > 0$ in our system. 
    Furthermore, Eqs.~\eqref{eq-theta0} and~\eqref{eq-chiN-sign} impose a constraint on the spectrum of the Hamiltonian 
    \begin{equation}
        e^{-i\lambda_s \tau} = e^{i\phi} (-1)^{N+s}.
    \end{equation}
    This leads to the following eigenvalue spacing condition
    \begin{equation}
        (\lambda_{s+1} - \lambda_s) \tau = (2k_s + 1) \pi. \label{eq-spectral_condition}
    \end{equation}
    Here, $k_s$ is an arbitrary non-negative integer that can vary with the index $s$.
    Conversely, it can be verified that if the Hamiltonian is mirror symmetric and its eigenvalues satisfy the spacing condition, then PST is guaranteed. Therefore, we conclude that the mirror symmetry condition Eq.~\eqref{eq-mirror_condition} and the eigenvalue spacing condition Eq.~\eqref{eq-spectral_condition} together constitute the necessary and sufficient criteria for realizing PST~\cite{2010_PST_Kay}. 

    The preceding discussion primarily focused on the special case $\theta=0$ (PST case). When $\theta \ne 0$, the system undergoes FST. In particular, when $\theta=\pi/4$, the system evolves into a maximally entangled state between $\ket{1}$ and $\ket{N}$, providing an essential resource for remote quantum information processing~\cite{2024_Malekakhlagh_EnhancedPST}. Furthermore, Eq.~\eqref{eq-FST} reveals an additional phase parameter $\psi$ that controls the relative phase between $\ket{1}$ and $\ket{N}$. We now turn to the general case where $\theta \ne 0$ and $\psi \ne 0$. From Eq.~\eqref{eq-expansion} we find that the term $\sin{\theta} + e^{i\psi} \cos{\theta}\chi_N$ has unit modulus, namely 
    \begin{align}
        \chi_N^2 + 2\tan{\theta}\cos{\psi}\chi_N - 1 = 0,
    \end{align}
    whose solutions, $\chi_{N,1}$ and $\chi_{N,2}$, satisfy
    \begin{eqnarray}
        \chi_{N,1} + \chi_{N,2} &=& -2 \tan{\theta} \cos{\psi}, \label{eq-chi_N-1}\\
        \chi_{N,1} \cdot \chi_{N,2} &=& -1. \label{eq-chi_N-2}
    \end{eqnarray}
    Two special values of $\psi$, namely $\psi=0$ and $\psi=\pi/2$, are of particular noteworthy and will be discussed separately below.

    Case 1: When $\psi=\frac{\pi}{2}$, solving Eqs.~\eqref{eq-chi_N-1} and \eqref{eq-chi_N-2}, together with the sign alternation of $\chi_N(\lambda_s)$ and the positivity condition of the weights in Eq.~\eqref{eq-recurr_weight}, yields
    \begin{equation}
        \chi_N(\lambda_s) = (-1)^{N+s},
    \end{equation}
    indicating that the mirror symmetry Eq.~\eqref{eq-mirror_condition} is maintained in this case. However, the eigenvalue spacing condition Eq.~\eqref{eq-spectral_condition} is violated, because
    \begin{equation}
        e^{-i\lambda_s \tau} = e^{i\phi} (\sin{\theta} + i \cos{\theta}(-1)^{N+s}) \ne e^{i\phi}(-1)^{N+s}.
    \end{equation}
    In this case, a specific set of eigenvalues can be prescribed~\cite{2016_FST_Genest}, from which the Hamiltonian can be reconstructed through the inverse eigenvalue process~\cite{2006_iep_gladwell}. The resulting Hamiltonian naturally satisfies the mirror symmetry condition, which is a central focus of this work.

    Case 2: When $\psi=0$, the corresponding solutions are
    \begin{equation}
        \chi_{N,1} = \tan(\frac{\pi}{4} - \frac{\theta}{2}),\quad \chi_{N,2} = -\cot(\frac{\pi}{4} - \frac{\theta}{2}),
    \end{equation}
    clearly, Eq.~\eqref{eq-chiN-sign} is not satisfied, and consequently the mirror symmetry condition Eq.~\eqref{eq-mirror_condition} is broken. However, since both sides of Eq.~\eqref{eq-expansion} must have unit modulus, we find that 
    \begin{equation}
        e^{-i\lambda_s \tau} = e^{i\phi} (\sin{\theta} + \cos{\theta}\chi_N(\lambda_s)) = e^{i\phi} (-1)^{N+s}.
    \end{equation}
    Thus, the eigenvalue spacing condition is identical to that of PST. 

    In this case, the FST Hamiltonian $H_{\text{FST}}$ can be constructed by applying an isospectral deformation $U$ to the PST Hamiltonian $H_{\text{PST}}$, i.e., $H_{\text{FST}} = U H_{\text{PST}} U^\dagger$~\cite{2016_FST_Genest}. This transformation preserves the eigenvalue spectrum of $H_{\text{PST}}$, while modifying only the frequency and coupling terms at the central sites. This type of FST has recently been experimentally demonstrated on the superconducting quantum platform~\cite{gpst_wang_2025}. Table~\ref{tab-review} summarizes the representative experiments demonstrated on various hardware platforms with different experimental configurations.

\section{Construction of the dome model }\label{sec-dome}
    \begin{figure*}
        \centering
        \includegraphics[width=1.6\columnwidth]{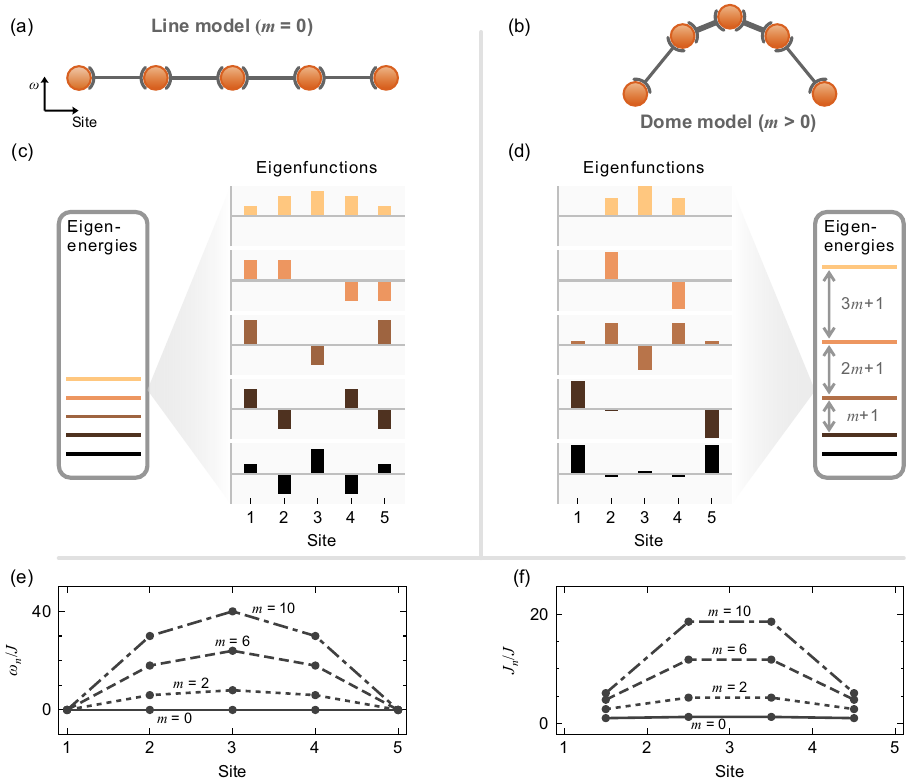}
        \caption{
            Eigenenergies, eigenfunctions, and Hamiltonian parameters of the line and dome model. 
            (a)--(b) Frequency and coupling configurations of the line and dome model, respectively. The vertical axis represents qubit frequencies, and the thickness of the gray lines indicates the relative coupling strength between adjacent qubits. 
            (c)--(d) Eigenenergies and corresponding eigenfunctions of the line and dome model, respectively. In the line model, all eigenenergies are equally spaced. In the dome model, all energy levels are pushed upward with increasing $m$, except for the two lowest levels, which remain unchanged. 
            (e)--(f) Hamiltonian parameters, including qubit frequencies $\omega_n$ and coupling strengths $J_n$, for various values of $m$. Parameters with $m=0$ correspond to the line model, while those with $m>0$ correspond to the dome model. 
        }
        \label{fig-fig2}
    \end{figure*}  

    \subsection{Physical Model}
Having established the spectral viewpoint in Sec.~\ref{sec-review}, we now turn to the construction of our new model.  In this section, we use the inverse eigenvalue method (IEM)~\cite{2006_iep_gladwell, 2005_iep_Chu} as a reconstruction tool  for designing Hamiltonians to realize FST and PST; a summary of the method is provided in Appendix~\ref{app:iem} . The IEM takes an eigenvalue spectrum as input, and we consider the following desired properties: 1. It should preferably generalize the eigenvalue spectrum of the line model, capable of reducing to it under specific conditions; 2. It should possess intrinsic noise resilience, potentially achieved by introducing specific energy gaps. 

    Based on these considerations, we propose the following general form for the eigenvalue spectrum 
    \begin{equation}
        \frac{\lambda_s(m)}{J} = s - \frac{N+1}{2} + (s-2)(s-1)\frac{m}{2}, 
        \label{eq-spectrum_FPST}
    \end{equation}
    where $s=1, 2, \cdots, N$. With this spectrum, the two lowest energy levels ($s = 1, 2$) remain unchanged, while the spacing between higher energy levels increases progressively. As $m$ increases, the low- and high-energy subspaces become effectively separated. When $m=0$, the spectrum reduces to the line model, thereby satisfying the two desired properties outlined at the beginning of this section. A schematic illustration of the resulting spectrum is shown in Fig.~\ref{fig-fig2}(d). Using the three-term recurrence relations summarized in Appendix~\ref{app:iem} , we iteratively compute the frequency and coupling terms of the Hamiltonian from Eqs.~\eqref{eq-recurr_wr} and \eqref{eq-recurr_Jr}, respectively, and obtain the eigenfunctions from Eqs.~\eqref{eq-ketlam} and \eqref{eq-vec_prin}. By analyzing the Hamiltonian structure across different $N$, we can generalize the expressions for the frequency and coupling terms that hold for arbitrary $N$
    \begin{equation}
        \omega_n = (n-1)(N-n)m J, \label{eq-dome_wn}
    \end{equation}
    and 
    \begin{eqnarray}
        J_n  =  && \frac{J}{2} \sqrt{n(N-n-1)m+n} \nonumber \\ 
        && \times \sqrt{(n-1)(N-n)m+N-n}.
        \label{eq-dome_Jn}
    \end{eqnarray}
    When $m = 0$, $J_n$ will be reduced to the conventional line model in Eq.~\eqref{eq-Jn}. Since the qubit frequencies in our model first increase and then decrease along the chain, forming a dome profile, we refer to this configuration as the dome model. 
    The dome model is valid when $m>0$, and $m$ takes values from $2, 6, 10,\cdots, 4k+2, \cdots$ ($k=0,1,2,\cdots$), under which both FST and PST can be realized. For $m=4k$, the system supports PST but not FST. For $m=2k+1$, the system can only undergo periodic evolution, returning to its initial state at $t=T$, and supports neither FST nor PST. The values of $\omega_n$ and $J_n$ under different $m$ are shown in Figs.~\ref{fig-fig2}(e) and \ref{fig-fig2}(f).

    The eigenvalues and eigenfunctions of the dome model exhibit distinct features compared to those of the line model. Unlike the equally spaced eigenenergies of the line model (Fig.~\ref{fig-fig2}(c)), the eigenenergies of the dome model (Fig.~\ref{fig-fig2}(d)) can be separated into two subspaces. 
    When $m$ is large, the low-energy subspace consists of only the two lowest eigenvalues, which remain unchanged with increasing $m$, while all other levels are pushed into a high-energy subspace that rises with $m$.
    The corresponding eigenfunctions also exhibit strong localization in the low-energy subspace, and the probability amplitudes are predominantly concentrated on the edge sites ($Q_1$ and $Q_5$). This subspace is well described by the effective SWAP Hamiltonian derived in Sec.~\ref{sec-swap}. 

    \subsection{Hidden SWAP dynamics in this Model \label{sec-swap}}
    \begin{figure}
        \centering
        \includegraphics{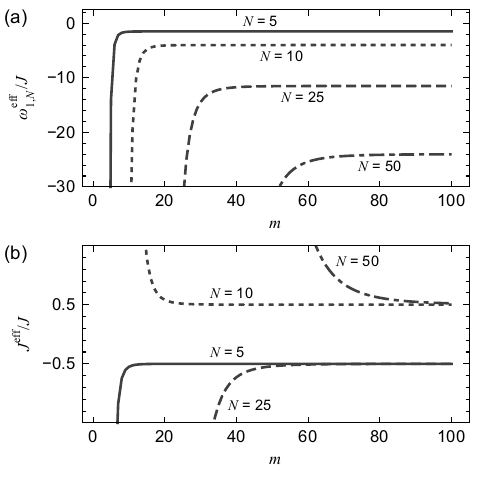}
        \caption{
        Numerical results of (a) the effective frequencies $\omega_{1, N}^\text{eff}$ and (b) the effective coupling strength $J^\text{eff}$ between the two end qubits in the large-$m$ limit. The calculation is based on a perturbative Schrieffer-Wolff transformation method, which block-diagonalizes the full $N \times N$ single-excitation Hamiltonian into a $2\times2$ subspace involving sites 1 and $N$, and an $(N-2)\times(N-2)$ subspace involving the remaining sites. This approximation holds only when $m$ is sufficiently large; for small $m$, the perturbative treatment breaks down, leading to divergent numerical results, which are not shown in the figure. 
        }
        \label{fig-fig3}
    \end{figure}

    The behavior of the dome configuration in the limiting cases of $m$ warrants detailed analysis. When $m=0$, the dome configuration naturally reduces to the conventional line model, as shown in Figs.~\ref{fig-fig2}(a) and \ref{fig-fig2}(c), where the coupling strengths follow Eq.~\eqref{eq-Jn} and all qubit frequencies are resonant.

Although the large-$m$ limit may not be directly accessible in experiments, it reveals an important hidden structure of the model. By analytically solving the effective Hamiltonian for small-$N$ cases and then extending the results by induction, one finds that, after integrating out the intermediate sites $n=2,3,\cdots,N-1$, the dynamics in the $m\to\infty$ limit reduce to an effective two-qubit Hamiltonian acting only on the edge sites~\cite{hidden_hu_2024}. The detailed derivation for the representative cases $N=3, 4$, and $5$ is presented in Appendix~\ref{app:swap}, while the general result is further checked numerically using a Schrieffer--Wolff transformation~\cite{swt_magesan_2020}, as shown in Fig.~\ref{fig-fig3}. The resulting effective parameters are 
    \begin{align}
        \lim_{m\rightarrow \infty}\omega_{1,N}^\text{eff} &= \omega_{1} - \frac{N-2}{2}J, \\
        \lim_{m\rightarrow \infty}J^\text{eff} &= (-1)^{N} \frac{J}{2} ,
    \label{eq-JefflargeN}
    \end{align}
    and therefore the effective Hamiltonian takes the form
    \begin{equation}
        H^{\text{eff}} = \begin{pmatrix}
            \omega_1^\text{eff} & J^{\text{eff}} \\
            J^{\text{eff}} & \omega_N^\text{eff}
        \end{pmatrix}. \label{eq-Heff}
    \end{equation}
    
    When $m$ is large, the intermediate qubits become strongly detuned, such that the condition $J \ll \abs{\Delta}$ is effectively satisfied. In this so-called dispersive regime~\cite{cqed_blais_2021}, real excitation exchange is suppressed, and the interaction between the edge qubits is mediated by virtual transitions through the intermediate sites. As a result, the intermediate sites are effectively integrated out, and an effective direct coupling between the edge qubits emerges. 
    In this sense, PST in the large-$m$ limit reduces to a hidden SWAP model between the left and right end sites~\cite{2019_coupler_krantz}. This effective description explains why the population on intermediate qubits is strongly suppressed and clarifies the physical origin of the enhanced robustness discussed below. 

    In summary, the dome model converges to two distinct effective models in different asymptotic regimes. For $m=0$, the $N$-qubit dome model becomes equivalent to the $N$-qubit line model. As $m \rightarrow \infty$, it reduces to a two-qubit SWAP model between the edge qubits.

\section{Dynamics of the dome model \label{sec-dynamic}}
    \begin{figure*}
        \centering
        \includegraphics[width=1.96\columnwidth]{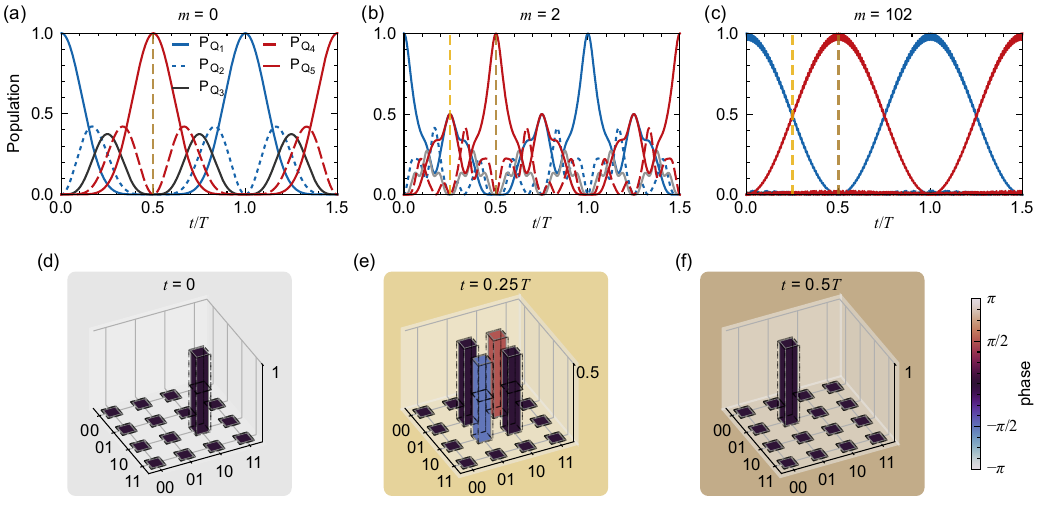}
        \caption{
            Qubit dynamics under different $m$ values and its reduction from PST to SWAP, simulated on a $1\times5$ qubit chain.  
            (a)--(c) Time evolution of qubit populations for $m=0$, 2, and 102, respectively. For $m\ne0$, both FST and PST occur sequentially at $t/T=0.25$ and 0.5, while for $m=0$, only PST is observed at $t/T=0.5$.  In case (c), when $m \gg 1$, the PST dynamics reduces to the SWAP dynamics between the two end sites, with effective coupling given by $J^\text{eff} = \pm J/2$; see Eq.~\eqref{eq-JefflargeN}. 
            (d)--(f) Density matrices of the two endpoint qubits $Q_1$ and $Q_5$ at $t/T=0$, 0.25, and 0.5, respectively. The bar height represents the amplitude of each element in the density matrix, while the color encodes its phase. 
            The simulations are performed under ideal and noise-free conditions, the density matrix in (d) corresponds to $t=0$ for all $m$ values; (e) corresponds to the result indicated by the light yellow dashed line in (b) and (c); and (f) corresponds to the result indicated by the dark yellow dashed line in (a)--(c).
        }
        \label{fig-fig4}
    \end{figure*}

    The dynamical properties of the dome model are noteworthy. Fig.~\ref{fig-fig4} shows simulations of the system dynamics for various values of $m$, using $N=5$ as an example. When $m=0$, the parameters provided by Eq.~\eqref{eq-dome_Jn} match those of the line model in Eq.~\eqref{eq-Jn}. Starting from the initial state $\ket{1}$, the system evolves to the last site $\ket{5}$ at half the period ($t/T=0.5$), returns to $\ket{1}$ at the full period ($t/T=1.0$), and continues to evolve periodically. Quantum state tomography~\cite{qst_2012_smolin} of the two end qubits reveals that at $t=0$ (Fig.~\ref{fig-fig4}(d)), the system is in the initial state $\rho_{Q_1Q_5}=\dyad{\uparrow\downarrow}{\uparrow\downarrow}$, while at $t/T=0.5$  (Fig.~\ref{fig-fig4}(f)) it evolves to $\rho_{Q_1Q_5}=\dyad{\downarrow\uparrow}{\downarrow\uparrow}$ with 100\% fidelity (assuming no coherent and decoherent errors), demonstrating a PST at this moment. 

    When $m=2$, the system enters the dome regime. In addition to PST at $t/T=0.5$, FST occurs at $t/T=0.25$, characterized by equal population exclusively on end qubits $\ket{1}$ and $\ket{5}$, with vanishing population on all other qubits, as indicated by the light yellow dashed line in Fig.~\ref{fig-fig4}(b). Quantum state tomography confirms that the system reaches a Bell entangled state $\rho_{Q_1Q_5}=\dyad{\psi}{\psi}$ between the two end qubits, with $\ket{\psi}=(\ket{\downarrow\uparrow} + i\ket{\uparrow\downarrow})/\sqrt{2}$ (Fig.~\ref{fig-fig4}(e)).  
    Our dome model is particularly noteworthy as it accomplishes both perfect state transfer and remote entanglement generation within a single period using the same set of Hamiltonian parameters. This feature not only provides a unified theoretical framework but also reduces experimental overhead. In contrast, in the line model, remote entanglement generation requires additional two-qubit gates beyond the PST Hamiltonian~\cite{2024_PST_XiangLiang}; meanwhile, schemes that generate remote entanglement directly via FST, such as the zig-zag model, do not share the same Hamiltonian for PST and FST, thereby requiring additional experimental calibration~\cite{gpst_wang_2025}. 

    As $m$ increases, an additional prominent feature of the dome model emerges, in which the PST dynamics reduce to SWAP dynamics.  
    This limit has been analytically and numerically discussed in subsection \ref{sec-swap}.  For example, at $m=102$, as shown in Fig.~\ref{fig-fig4}(c), the system retains all features observed at $m=2$ while significantly suppressing the population on intermediate qubits ($Q_2, Q_3$, and $Q_4$). This suggests that the dome model at large-$m$ exhibits enhanced noise resilience, a point that will be discussed in detail in the following sections. For larger $N$, the Hamiltonian remains analytically solvable using Eq.~\eqref{eq-dome_Jn}, and its dynamics exhibit the same key features. 

\section{Noise resilience of the dome model \label{sec-noise}}
    \begin{figure*}
        \centering
        \includegraphics[width=1.88\columnwidth]{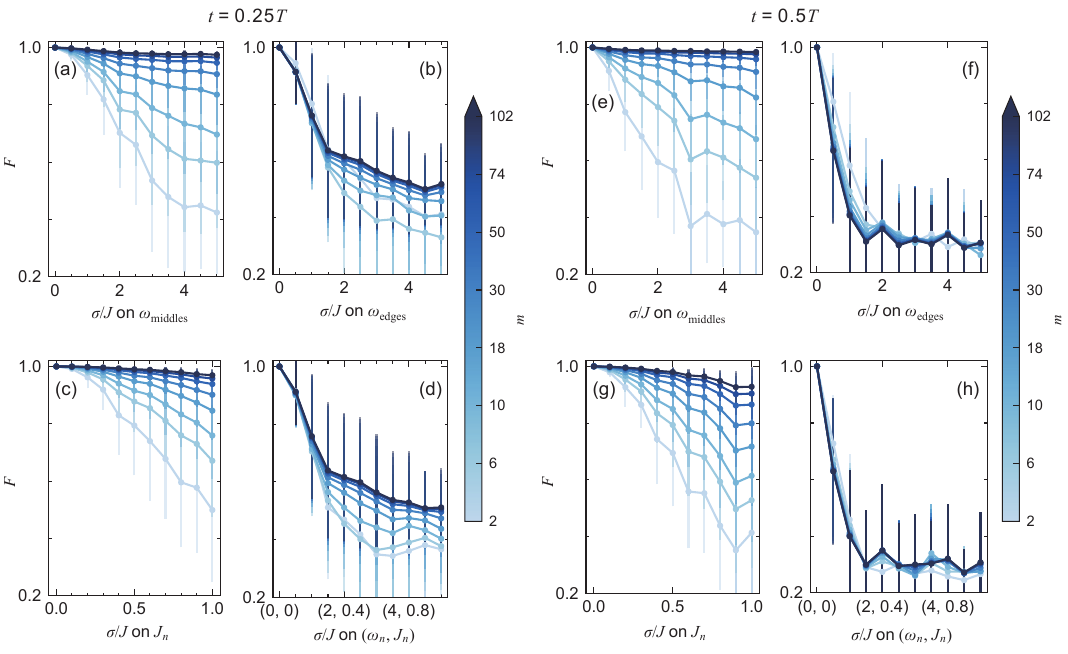}
        \caption{
            Impact of different types of coherent noise on the fidelities of remote entanglement generation and quantum state transfer, simulated on a $1\times5$ qubit chain. 
            (a)--(d) Fidelities of entanglement generation at $t=0.25T$ under various noise types. (a) Noise on the middle qubit frequencies $\omega_{\text{middles}}$ ($Q_2$--$Q_4$); (b) Noise on the edge qubit frequencies $\omega_{\text{edges}}$ ($Q_1$ and $Q_5$); (c) Noise on the coupling strengths $J_n$ ($J_1$--$J_4$); (d) Noise on all components. Noise is sampled from a Gaussian distribution centered at the ideal parameter values with a variance of $\sigma^2$. Each data point represents the average fidelity over 100 samples, with the error bars indicating the standard deviation. The fidelity of entanglement generation is defined as the overlap between the density matrix $\rho$ of the two edge qubits at $t=0.25T$ and the ideal density matrix $\dyad{\psi}{\psi}$, i.e., $F(t=0.25T) = \mathrm{Tr}(\rho \dyad{\psi}{\psi})$, where $\ket{\psi} = (\ket{\downarrow\uparrow} + i\ket{\uparrow\downarrow})/\sqrt{2}$ is a Bell entangled state.
            (e)--(h) Fidelities of quantum state transfer at $t=0.5T$ under the same noise types as (a)--(d), respectively. Each data point is averaged over 50 samples. Fidelity is defined as the overlap between the reconstructed process matrix $\chi$, obtained via the quantum process tomography simulation, and the ideal matrix $\chi_{\text{ideal}}$, i.e., $F(t=0.5T) = \mathrm{Tr}(\chi \chi_{\text{ideal}})$. Details of the reconstruction of $\chi_{\text{ideal}}$ are provided in the main text.
        }
        \label{fig-fig5}
    \end{figure*}

    \begin{figure*}
        \centering
        \includegraphics[width=1.9\columnwidth]{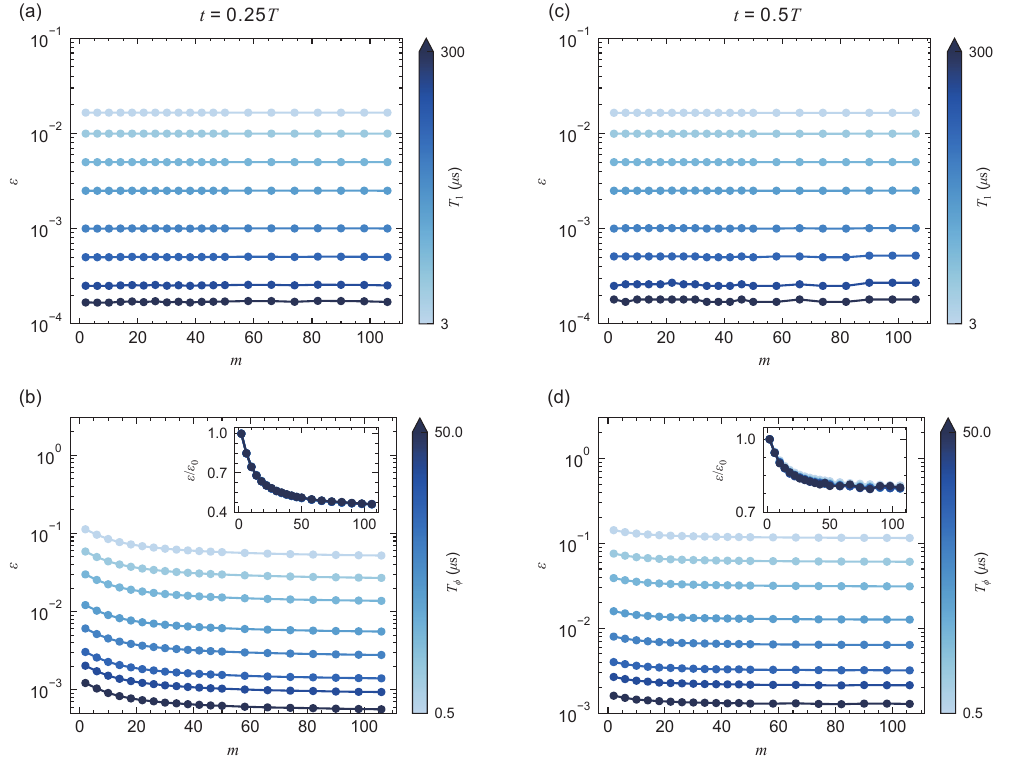}
        \caption{
            Impact of decoherence noise on the errors of remote entanglement generation and quantum state transfer, simulated on a $1\times5$ qubit chain, with an evolution period of $T=200$~ns.  
            (a) Error of entanglement generation at $t=0.25T$ as a function of $m$ and the relaxation time $T_1$, with dephasing neglected ($T_{\phi} \to \infty$). The value of $m$ increases from 2 to 102. The blue curves, from light to dark, correspond to $T_1 = [3, 5, 10, 20, 50, 100, 200, 300]$~$\mu$s. 
            (b) Error of entanglement generation as a function of $m$ and the dephasing time $T_{\phi}$, with relaxation neglected ($T_1 \to \infty$). The blue curves, from light to dark, correspond to $T_{\phi} = [0.5, 1, 2, 5, 10, 20, 30, 50]$~$\mu$s. Inset: normalized error $\varepsilon/\varepsilon_0$ as a function of $m$ and $T_{\phi}$, where $\varepsilon_0$ denotes the error at $m=0$. 
            (c), (d) Errors of quantum state transfer at $t=0.5T$ under the same conditions as in (a) and (b), respectively. Decoherence is applied uniformly to all qubits, assuming identical coherence times. The error is defined as $\varepsilon \equiv 1-F$, where the definitions of fidelity $F$ are provided in Sec.~\ref{sec-noise} of the main text. 
        }
        \label{fig-fig6}
    \end{figure*}
    
    \subsection{Coherent noise \label{subsec-coherent-noise}}
    Experimental imperfections, such as the parameter deviations due to calibration precision~\cite{2024_ParityPST_Roy, 2023_PST_ZhangChi, 2018_PST_LiX}, as well as drifts in qubit frequencies and coupling strengths caused by environmental instability~\cite{drift_2018_meifeng, drift_2019_google, drift_2024_mcewen, 2020_tls_graaf, 2022_tls_bejanin, 2023_tls_thorbeck, perturb_ronke_2011}, can be categorized as coherent errors or parameter disorder. To investigate the impact of coherent noise on fidelity, we intentionally introduce perturbations to the ideal parameters and simulate the fidelity variations. Specifically, we generate noise samples from a Gaussian distribution $\mathcal N(0, \sigma^2)$ using the NumPy random number generator and add them to the ideal parameters, yielding the perturbed parameters
    \begin{eqnarray}
        && J_n \rightarrow J_n + \xi_n, \quad \xi_n \sim \mathcal N(0, \sigma^2),  \\ 
        && \omega_n \rightarrow \omega_n + \eta_n, \quad \eta_n \sim \mathcal N(0, \sigma^2). 
    \end{eqnarray}
    Fig.~\ref{fig-fig5} shows simulations of a $1\times5$ qubit chain, where we evaluate the effects of noise on four types of parameters: 1. Middle qubit ($Q_2$ -- $Q_4$) frequencies $\omega_{\text{middles}}$; 2. Edge qubit ($Q_1$ and $Q_5$) frequencies $\omega_{\text{edges}}$; 3. Coupling strengths $J_n$ between $Q_n$ and $Q_{n+1}$; and 4. All components combined, including both $\omega_n$ and $J_n$. 

    Before analyzing the simulation results, we clarify the specific definitions of fidelity used in different scenarios. Panels (a)--(d) in Fig.~\ref{fig-fig5} represent the fidelity of entanglement generation at $t=0.25T$, defined as $F(t=0.25T) = \mathrm{Tr}(\rho_{Q_1Q_5} \dyad{\psi}{\psi})$. Here $\rho_{Q_1Q_5}$ is the simulated density matrix of two edge qubits at $t=0.25T$, and $\ket{\psi}$ is the Bell entangled state defined earlier. Panels (e)--(h) in Fig.~\ref{fig-fig5} display the fidelity of quantum state transfer at $t=0.5T$, defined as $F(t=0.5T) = \mathrm{Tr}(\chi \chi_{\text{ideal}})$. The process matrix $\chi$ is reconstructed as follows. At $t=0$, four initial states, $\{ \ket{\downarrow}, \ket{\uparrow}, (\ket{\downarrow}+\ket{\uparrow})/\sqrt{2}, (\ket{\downarrow}+i\ket{\uparrow})/\sqrt{2} \}$, are prepared on the first qubit $Q_1$. After evolution to $t=0.5T$, three single-qubit gates, $\{ \mathbb{I}, X_{\pi/2}, Y_{\pi/2} \}$, are applied to the final qubit $Q_5$ to perform measurements in different Pauli bases. $\chi$ is then reconstructed using quantum process tomography~\cite{qpt_2013_korotkov, qpt_2013_merkel, qpt_2018_knee}. $\chi_{\text{ideal}}$ represents the ideal quantum state transfer process, which, from the perspective of $Q_5$, corresponds to an identity operation~\cite{2018_PST_LiX}.

    The simulation results in Figs.~\ref{fig-fig5}(a) and \ref{fig-fig5}(c) (\ref{fig-fig5}(e) and \ref{fig-fig5}(g)) demonstrate that as $m$ increases, the fidelity of entanglement generation (quantum state transfer) becomes increasingly robust against noise on middle qubit frequencies $\omega_{\text{middles}}$ and coupling strengths $J_n$. 
    However, as shown in Figs.~\ref{fig-fig5}(b) and \ref{fig-fig5}(f), the fidelities show no significant improvement against noise on edge qubit frequencies $\omega_{\text{edges}}$ with increasing $m$. 
    This phenomenon can be understood based on the analysis in Sec.~\ref{sec-dome}. When $m$ is sufficiently large, all intermediate sites are effectively integrated out, resulting in an effective SWAP Hamiltonian between the edge sites. As a result, increasing $m$ enhances the robustness against noise on all intermediate components, including $\omega_{\text{middles}}$ and $J_n$. In contrast, the effect of noise on $\omega_{\text{edges}}$ remains significant, since the edge sites are retained in the effective model.
    The results in Figs.~\ref{fig-fig5}(d) and \ref{fig-fig5}(h) indicate that when noise is present on all components, the fidelity is ultimately limited by the weakest link, namely, the noise on $\omega_{\text{edges}}$. Since coherent errors can be mitigated through careful experimental control (e.g., placing superconducting transmon qubits at their sweet spot to suppress frequency fluctuations~\cite{2019_coupler_krantz}), the results in Fig.~\ref{fig-fig5} highlight the importance of minimizing such errors on edge qubits in experiments. 
    
    \subsection{Decoherence noise \label{subsec-decoherent-noise}}
    In addition to coherent errors, the system also experiences decoherence during experiments~\cite{2019_coupler_krantz, deco_2019_burnett, deco_2006_caijianming}. This work mainly focuses on the relaxation and dephasing noise, quantified by the relaxation time $T_1$ and pure dephasing time $T_{\phi}$~\cite{2019_coupler_krantz}, respectively. 
    To ensure meaningful experimental guidance, we simulate decoherence using parameters that match the performance of state-of-the-art quantum platforms. 
    In the simulation presented in Fig.~\ref{fig-fig6}, we fix the evolution rate at $J/2\pi = 5$~MHz (corresponding to a period of $T = 200$~ns), with $T_1$ scanned from 3 to 300~$\mu$s and $T_{\phi}$ scanned from 0.5 to 50~$\mu$s. The wide range of $T_1$ and $T_{\phi}$ spans nearly all current superconducting quantum platforms~\cite{2025_newest_willow, 2025_newest_Zuchongzhi3.0, 2025_newest_FanHeng, 2024_newest_IBM, 2024_newest_baqis, 2024_PST_XiangLiang}, ensuring the generality of the results. We expect the conclusion drawn from Fig.~\ref{fig-fig6} to generalize to other quantum platforms. 

    Figs.~\ref{fig-fig6}(a) and \ref{fig-fig6}(c) show that, as $T_1$ increases from $3~\mu$s to $300~\mu$s, the error $\varepsilon \equiv 1-F$ decreases by nearly two orders of magnitude. For example, in Fig.~\ref{fig-fig6}(a) at $m=0$, it decreases from $1.65\times10^{-2}$ to $1.67\times10^{-4}$. The error therefore follows the scaling relation $\varepsilon \propto 1/T_1$, consistent with the prediction of standard decoherence models~\cite{deco_omalley_2015, deco_abad_2022, deco_abad_2025, deco_ding_2023, deco_li_2024, deco_marxer_2023, deco_marxer_2026}. 
    Similarly, Figs.~\ref{fig-fig6}(b) and ~\ref{fig-fig6}(d) show that, as $T_{\phi}$ increases from $0.5~\mu$s to $50~\mu$s, the error decreases by nearly two orders of magnitude, following the scaling relation $\varepsilon \propto 1/T_{\phi}$. For example, in Fig.~\ref{fig-fig6}(b) at $m=0$, it decreases from $1.13\times10^{-1}$ to $1.22\times10^{-3}$.
    
    However, as shown in Figs.~\ref{fig-fig6}(a) and \ref{fig-fig6}(c), we do not observe any significant dependence of the error on $m$ at any $T_1$ value. We attribute this behavior to the fact that while increasing $m$ suppresses the population on the intermediate qubits during the evolution (see Fig.~\ref{fig-fig4}), the population on the edge qubits increases correspondingly in order to conserve the total excitation number. This redistribution of population (more precisely, probability amplitude) does not change the cumulative effect of relaxation. 

    In contrast, for large $m$, the dynamics predominantly involve coherent superpositions of the edge-qubit states. As a result, the originally global dephasing process becomes effectively localized to the two edge qubits, substantially reducing the impact of dephasing noise. As shown in the inset of Fig.~\ref{fig-fig6}(b), for sufficiently large $m$, the normalized error $\varepsilon/\varepsilon_0$ is reduced to approximately 40~\% of that at $m=0$. This is consistent with the expectation from an uncorrelated $N$-qubit noise model, in which the effective shutdown of the $(N-2)$ noise channels associated with the intermediate sites reduces the error by a factor of approximately $2/N$~\cite{deco_abad_2022}.

    However, as shown in the inset of Fig.~\ref{fig-fig6}(d), the reduction in $\varepsilon / \varepsilon_0$ is less pronounced than that observed in the inset of Fig.~\ref{fig-fig6}(b). 
    This difference originates from the definition of the error. While Fig.~\ref{fig-fig6}(b) is based on quantum-state fidelity, Fig.~\ref{fig-fig6}(d) measures the process error, i.e., $\varepsilon(t=0.5T)=1-\mathrm{Tr}(\chi\chi_{\mathrm{ideal}})$. 
    As discussed in Sec.~\ref{subsec-coherent-noise}, reconstructing $\chi$ requires four input states, three of which contain a ground-state component. 
    Compared with the single-excitation component initially on $Q_1$, which spreads over the entire single-excitation subspace during the evolution, the ground-state component remains stationary throughout the evolution and is therefore less affected by dephasing. Consequently, the process fidelity shows a weaker dependence on $m$ than the state fidelity shown in Fig.~\ref{fig-fig6}(b).

    We emphasize that the above discussion is only a semi-quantitative interpretation based on a first-order dissipative approximation and an uncorrelated noise model~\cite{deco_abad_2022, deco_abad_2025}. A more complete understanding of multiqubit dynamics in open quantum systems remains an interesting direction for future investigation~\cite{2010_PST_Kay, deco_2006_caijianming, manybody_2013_pingy, manybody_2020_zhangshunyao, manybody_2023_busel}.

\section{Scalability of the dome model \label{sec-scale}}

    \begin{figure}
        \centering
        \includegraphics[width=0.92\columnwidth]{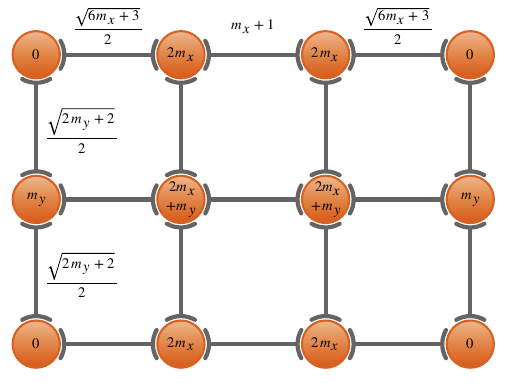}
        \caption{Parameters of the 2D $3\times 4$ qubit network for FST and PST.}
        \label{fig-fig7}
    \end{figure}

    \begin{figure*}
        \centering
        \includegraphics[width=1.9\columnwidth]{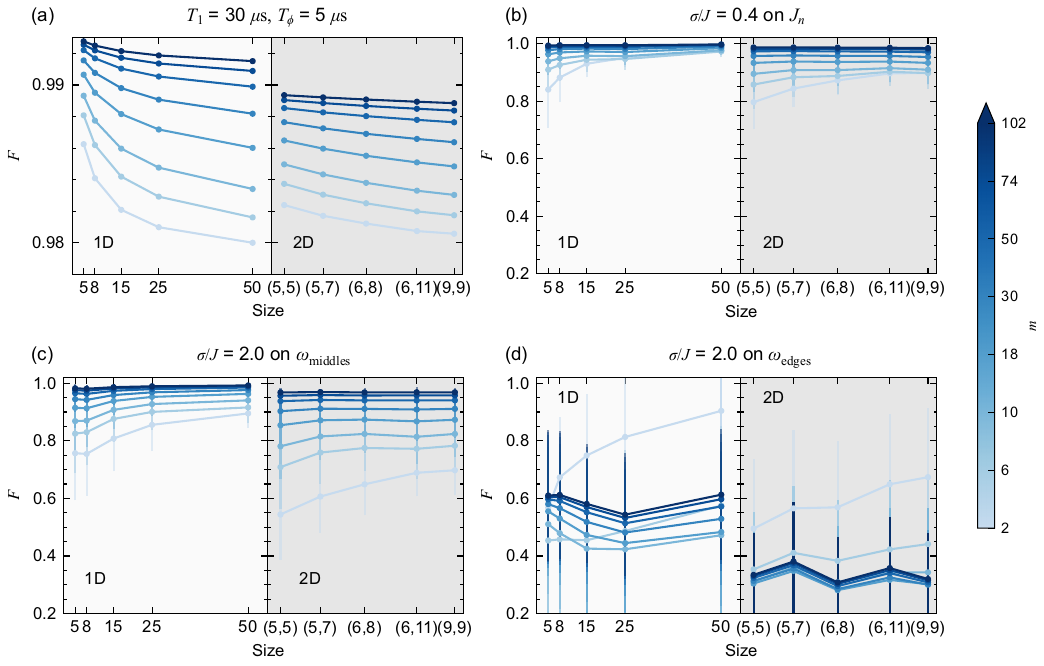}
        \caption{
            Fidelity of remote entanglement generation under different system sizes and noise types. 
            (a) Fidelity as a function of system size, with relaxation time $T_1$ fixed at 30~$\mu$s, dephasing time $T_{\phi}$ fixed at 5~$\mu$s, and evolution period $T=200$~ns. Coherent noise is not considered. The left panel shows results for 1D qubit chains of length $N$, while the right panel corresponds to 2D qubit networks of size $(R, C)$, where $R$ and $C$ denote the number of rows and columns, respectively. The $(R, C)$ labels on the x-axis are positioned according to $\sqrt{R \times C}$. 
            (b) Fidelity versus system size with Gaussian noise ($\sigma/J=0.4$) applied to the coupling strengths $J_n$, excluding decoherence. 
            (c)--(d) Fidelity versus system size with Gaussian noise ($\sigma/J=2.0$) applied to (c) the middle qubit frequencies $\omega_{\text{middles}}$ and (d) the edge qubit frequencies $\omega_{\text{edges}}$, excluding decoherence. For 1D chains, the edge qubits are $Q_1$ and $Q_N$; for 2D networks, they are $Q_{(1, 1)}, Q_{(1, C)}, Q_{(R, 1)},$ and $Q_{(R, C)}$. The middle qubits comprise all non-edge qubits. 
            Fidelity is computed as $F(t=0.25T) = \mathrm{Tr}(\rho \rho_{\text{ideal}})$, where $\rho_{\text{ideal}}$ denotes the ideal reduced density matrix of the edge qubits at $t=0.25T$. 
            In the 1D case, $\rho_{\text{ideal}}=\dyad{\psi}{\psi}$, with $\ket{\psi} = (\ket{\downarrow\uparrow} + i\ket{\uparrow\downarrow})/\sqrt{2}$ a Bell entangled state. In the 2D case, $\rho_{\text{ideal}}=\dyad{\mathcal{W}}{\mathcal{W}}$, with $\ket{\mathcal{W}} = (\ket{\uparrow\downarrow\downarrow\downarrow} + i\ket{\downarrow\uparrow\downarrow\downarrow} + i\ket{\downarrow\downarrow\uparrow\downarrow} + \ket{\downarrow\downarrow\downarrow\uparrow})/2$ a $\mathcal{W}$ entangled state. 
            Each data point in (b)--(d) represents the average fidelity, with error bars indicating the standard deviations over 50 random samples. 
        }
        \label{fig-fig8}
    \end{figure*}

    All simulation results presented so far are demonstrated on a $1\times5$ qubit chain. However, Table~\ref{tab-review} shows that current experimental implementations already exceed this scale, reaching up to $1\times11$ in 1D and $6\times6$ in 2D architectures. Therefore, it is necessary to discuss the scalability of the dome model. Extension to a 1D qubit chain with arbitrary length $N$ is straightforward, as the Hamiltonian parameters $\omega_n$ and $J_n$ are explicitly given by Eqs.~\eqref{eq-dome_wn} and \eqref{eq-dome_Jn}, respectively. Here, we focus on the extension of the dome model to the 2D case. As discussed in Refs.~\onlinecite{2010_PST_Kay, 2024_PST_XiangLiang}, it is essential that the Hamiltonian parameters along both the $x$ and $y$ dimensions independently satisfy the conditions Eqs.~\eqref{eq-dome_wn} and \eqref{eq-dome_Jn} while maintaining the same evolution period. As a concrete example shown in Fig.~\ref{fig-fig7}, we consider a $3\times4$ qubit network implementing the dome model. According to Eq.~\eqref{eq-dome_Jn}, the coupling strengths in the $x$ direction (i.e., for each row) must satisfy
    \begin{equation}
        J_n^x / J = \{ \frac{\sqrt{6m_x+3}}{2}, m_x+1, \frac{\sqrt{6m_x+3}}{2} \},
    \end{equation}
    while those along the $y$ direction (i.e., for each column) must satisfy 
    \begin{equation}
        J_n^y / J = \{ \frac{\sqrt{2m_y+2}}{2}, \frac{\sqrt{2m_y+2}}{2} \}.
    \end{equation}
    Meanwhile, the frequency parameters in the $x$ and $y$ directions should satisfy 
    \begin{equation}
       \omega_n^x / J = \{ 0, 2m_x, 2m_x, 0 \},\quad \omega_n^y / J = \{ 0, m_y, 0 \}.
    \end{equation}
    The structure of this model is presented in Fig. \ref{fig-fig7}. 

    Since only the relative frequencies within each set influence the dynamics, a global offset does not affect the results. 
    Therefore, to ensure that the frequencies in each row and column satisfy the conditions in Eq.~\eqref{eq-dome_wn}, the frequencies for the 1st, 2nd and 3rd rows are set as $\{ 0, 2m_x, 2m_x, 0 \} \times J$, $\{ m_y, 2m_x + m_y, 2m_x + m_y, m_y \} \times J$, $\{ 0, 2m_x, 2m_x, 0 \} \times J$, respectively.
    In the dome model, $m_x$ and $m_y$ must take values from $2, 6, 10, \cdots$. Under the dome regime in a 2D architecture, a state initialized in $\ket{1}$ can, at $t=0.25T$, evolve fractionally into a maximally entangled $\mathcal{W}$ state among the four corner qubits, i.e., $\ket{\mathcal{W}} = (\ket{\uparrow\downarrow\downarrow\downarrow} + i\ket{\downarrow\uparrow\downarrow\downarrow} + i\ket{\downarrow\downarrow\uparrow\downarrow} + \ket{\downarrow\downarrow\downarrow\uparrow})/2$. When $m_x=m_y=0$, the model reduces to the 2D line model, which supports PST but, unlike the dome model, does not allow FST for entanglement generation. 

    Fig.~\ref{fig-fig8} illustrates how fidelities under different types of noise change with system size. As Figs.~\ref{fig-fig5} and \ref{fig-fig6} show that our conclusions hold for both entanglement generation and quantum state transfer, Fig.~\ref{fig-fig8} presents only the fidelities of entanglement generation. 
    Fig.~\ref{fig-fig8}(a) shows results with only decoherence noise considered (with $T_1=30~\mu$s and $T_{\phi}=5~\mu$s for all qubits), indicating that increasing $m$ improves fidelity regardless of the system size. 
    Figs.~\ref{fig-fig8}(b) and \ref{fig-fig8}(c) indicate that increasing $m$ enhances the robustness against noise on both $J_n$ and $\omega_{\text{middles}}$, across all system sizes. The overall fidelity tends to increase with the system size. This is because under a fixed evolution period, a larger system size corresponds to greater parameter magnitudes, making a given noise level cause relatively smaller perturbations. 
    Fig.~\ref{fig-fig8}(d) shows that increasing $m$ does not significantly improve fidelity under the noise on $\omega_{\text{edges}}$. This agrees with the observations in Figs.~\ref{fig-fig5}(b) and \ref{fig-fig5}(f), where we concluded that increasing $m$ effectively integrates out the middle qubits but retains the edge qubits, thus failing to enhance resilience to noise on $\omega_{\text{edges}}$. 

\section{Long-distance PST \label{sec-longpst}}
    Finally, we discuss the limitations of these PST and FST schemes, including the one proposed in this work, when scaled to much larger physical systems, and suggest possible ways to overcome these limitations in experimental implementations. To date, the largest PST demonstrated in experiment involves a $6\times6$ system, as summarized in Table~\ref{tab-review}. 
    The challenge in much larger systems lies in the limitations of maximum achievable coupling. When $N$ is large, we have
    \begin{align} 
        J_\text{max} &= \frac{N J}{4}, \quad \text{for line model}, \label{eq-line_ratio}\\
        J_\text{max} &= \frac{mN^2J}{8}, \quad \text{for dome model}. \label{eq-dome_ratio}
    \end{align}
    For both the line and dome models, the maximum coupling strength appears at the central sites. In practical experimental devices, there exists an upper bound on the coupling strength, so increasing $N$ leads to reduced $J$ values, resulting in longer transfer time and making the system more susceptible to decoherence~\cite{2019_coupler_krantz, deco_2019_burnett, deco_2006_caijianming}. From another perspective, by considering the hardware-imposed upper bound on $J_\text{max}$ and the decoherence-limited lower bound on $J$, one can estimate the maximum achievable system size $N$. For instance, assuming $J_\text{max}/2\pi \le 50$~MHz and $J/2\pi \ge 0.5$~MHz, we estimate that the line model is limited to $N = 4J_\text{max} / J \le 400$. In contrast, for the dome model with $m = 10$, the bound becomes $N = \sqrt{8J_\text{max} / mJ} \le 9$, posing a severe challenge to scalability. 

\begin{figure}
    \centering \includegraphics[width=0.45\textwidth]{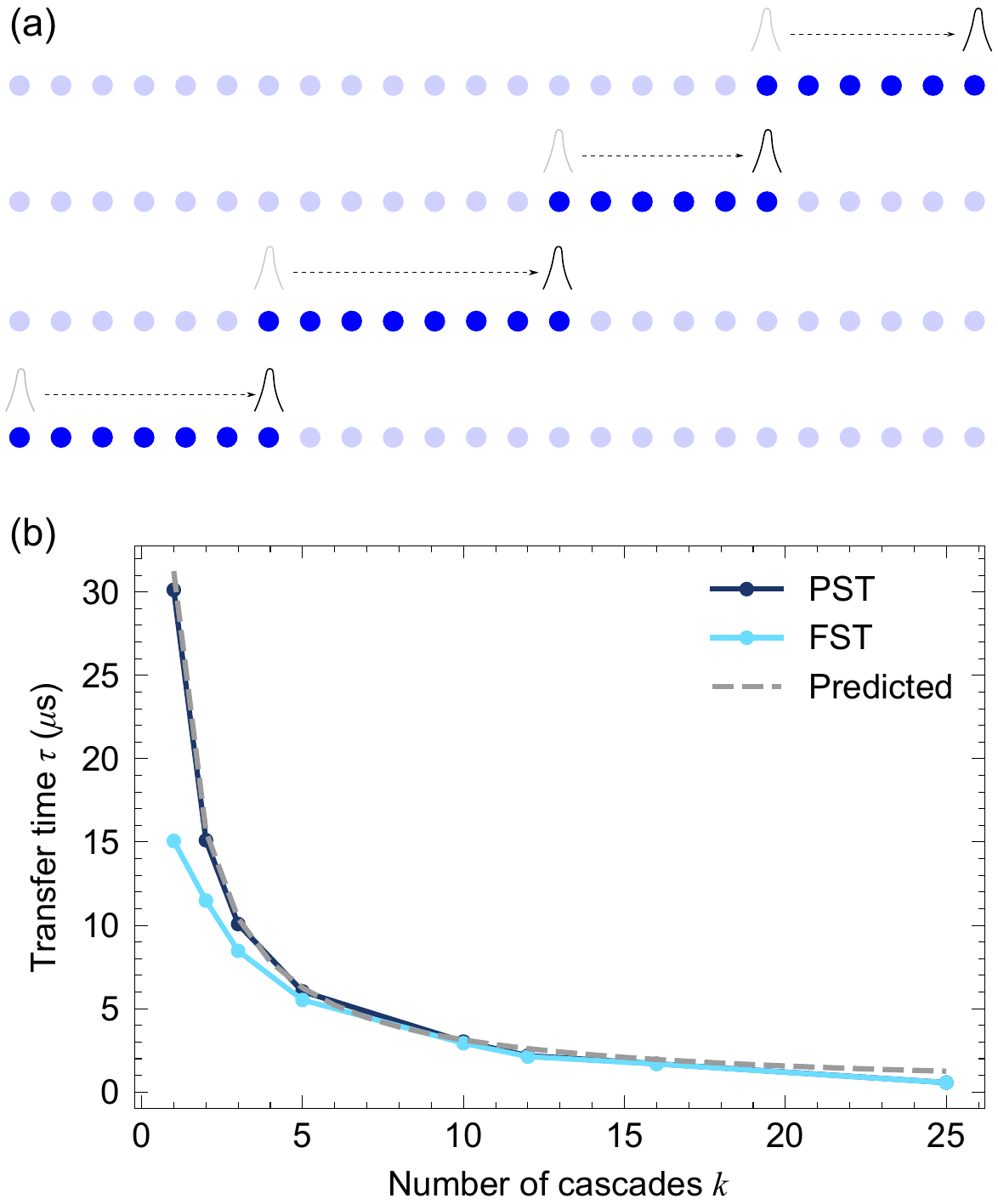}
    \caption{
    Long-distance state transfer via the cascading scheme. 
    (a) Schematic illustration of decomposing a long-distance PST into a sequence of sub-PST steps. 
    (b) Total transfer time $\tau$ as a function of the number of cascades $k$. In the simulation we set $N=50, m=10$, and $J_\text{max}/2\pi=50$~MHz. The subchain lengths are chosen to be as uniform as possible, and the shared-site overhead is taken into account. For example, when $k=3$, the chain is divided into subchains of lengths 17, 17, and 18. The dashed line shows the estimated total transfer time based on Eq.~\eqref{eq-tau_dome}, in excellent agreement with the numerical calculations. 
    }
    \label{fig-fig9}
    \end{figure}

    To address these issues, an intuitive idea is to split the whole system into many subsystems, and a single PST process is divided into $k$ sequential executions~\cite{cascade_alsulami_2022, cascade_alsulami_2024} 
    \begin{equation}
    |1\rangle \rightarrow |N_1\rangle \rightarrow |N_2\rangle  \rightarrow |N_3 \rangle \rightarrow \cdots \rightarrow |N_k = N\rangle,
    \end{equation}
    as schematically shown in Fig.~\ref{fig-fig9}(a). 
    The reduced parameter space in each sub-PST further simplifies the parameter calibration and optimization. Moreover, this cascaded proposal is also experimentally feasible. For instance, in superconducting transmon qubit systems equipped with tunable couplers~\cite{2023_coupler_ZhangChi, 2019_coupler_krantz, 2018_coupler_yanfei}, the coupling strength can be tuned from 0 to approximately 50~MHz, with switching times below 5~ns. This is much shorter than the typical PST transfer time of 25 to 400~ns listed in Table~\ref{tab-review}, indicating that cascading two PST processes would incur almost no additional time overhead. 

    A straightforward calculation shows that in the line model, the total duration of the cascaded proposal is not shorter than that of a single PST. For simplicity, we assume that each sub-PST spans $N/k$ qubits. In this case, the evolution rate for each sub-PST is calculated from Eq.~\eqref{eq-Jn} as
    \begin{equation}
        J_\text{sub} = \frac{4J_\text{max}}{N/k},
    \end{equation}
    the total transfer time is given by 
    \begin{equation}
        \tau = k \frac{\pi}{J_\text{sub}} = \frac{\pi N}{4 J_\text{max}}. \label{eq-tau_line}
    \end{equation}
    Here, we assume $N \gg k$ and neglect the shared site between two consecutive sub-PST processes.
    Eq.~\eqref{eq-tau_line} indicates that, for the line model, splitting PST into multiple segments does not yield any reduction in the total transfer time. However, when $k$ is comparable to $N$, the shared-site overhead can even make the total time exceed that of a single PST. For example, splitting PST into $(N-1)$ sequential SWAP operations leads to a total transfer time of $\tau = \pi (N-1) / 2J_\text{max}$, nearly doubling the duration~\cite{speed_yung_2006, 2024_PST_XiangLiang}. This behavior, however, will be totally changed in the dome model. 

    For the dome model, Eq.~\eqref{eq-dome_Jn} shows that for each sub-PST, when $N$ and $m$ are large, the evolution rate is given by 
    \begin{equation}
        J_\text{sub} = \frac{8J_\text{max}}{m (N/k)^2},
    \end{equation}
    when $N \gg k$, the total transfer time of $k$ cascaded sub-PSTs is 
    \begin{equation}
        \tau = k \frac{\pi}{J_\text{sub}} = \frac{\pi m N^2}{8kJ_\text{max}}. \label{eq-tau_dome}
    \end{equation}
    Eq.~\eqref{eq-tau_dome} indicates that, by splitting a single PST into a cascade of $k$ sub-PSTs, the transfer time can be significantly reduced by a factor of $k$. 
    As an example, Fig.~\ref{fig-fig9}(b) shows that for PST in a $1\times50$ qubit chain with $m=10$, the transfer time without cascading reaches $\tau \approx 30~\mu$s, which is comparable to or even exceeds the coherence time of current superconducting quantum processors. By contrast, setting $k=10$ reduces the total transfer time to $\tau \approx 3~\mu$s, bringing it well within an experimentally accessible regime and thereby significantly enhancing the practical feasibility of the dome model.
    Therefore, to highlight the advantages of the dome model in experiments, one must balance the choice of $k$ and $m$; the former determines the total transfer time and thus affects the decoherence effect, and the latter determines the noise-suppression capability of each sub-PST. With the cascaded proposal, the dome model is expected to play an important role in long-distance PST and FST. Furthermore, we anticipate that long-distance PST could be demonstrated on other hardware platforms, such as optical waveguides~\cite{2016_optical_Chapman}. 

    With the cascaded proposal for PST as discussed above, we now briefly outline the corresponding strategy for FST. A single FST process cannot be directly divided into multiple sub-FSTs, as this would fail to generate the desired Bell state between the two end qubits. Therefore, the feasible approach is to split a single FST into one sub-FST and $(k-1)$ sub-PSTs, with a total duration still approximately given by Eq.~\eqref{eq-tau_dome} when $k$ is large, as shown in Fig~\ref{fig-fig9}(b). In comparison, the duration of a single FST is $\tau = \pi m N^2 / 16J_\text{max}$, and in this way for $k > 2$ the cascaded proposal still provides a clear advantage. Furthermore, the cascaded proposal allows for techniques such as quantum state purification~\cite{purify_yan_2022} or error-correcting codes~\cite{qec_burkhart_2021} to be applied in each sub-process, thereby improving the fidelity of each segment and ultimately enhancing the overall performance of both FST and PST. 

\section{Conclusion \label{sec-conclusion}}
    In conclusion, we summarize a methodology for solving the inverse eigenvalue problem~\cite{2006_iep_gladwell, 2005_iep_Chu}, which requires only a specified eigenvalue spectrum and the constraint that the Hamiltonian be tridiagonal and mirror symmetric. Using this approach, we construct the dome model, whose parameters can be fully solved analytically for arbitrary scales in both 1D and 2D cases. This model enables the sequential realization of remote entanglement generation (at 1/4 period) and perfect state transfer (at 1/2 period) within a single evolution cycle. In 1D qubit chains, the entanglement manifests as Bell states between the two end qubits, whereas in 2D qubit networks, it corresponds to $\mathcal{W}$ states among the four corner qubits.  
    The model proposed in this work introduces a tunable parameter $m$ associated with the energy gap between adjacent sites and gives rise to an elegant underlying structure. Specifically, when $m=0$, the dome model naturally reduces to the conventional line model. In the large-$m$ limit, the intermediate sites are effectively integrated out, yielding an effective SWAP model involving only the two edge qubits and thereby making the fidelity nearly immune to noise on intermediate sites. Finally, we propose a cascaded scheme that enables the experimental realization of long-distance FST and PST.

    We have also employed this methodology to construct other types of Hamiltonians. For instance, our proposed zig-zag model not only achieves perfect state transfer but also exhibits a certain noise suppression effect~\cite{gpst_wang_2025}. Furthermore, by applying an isospectral transformation to the Hamiltonian, this model can also facilitate remote entanglement generation. These unique properties of the zig-zag model have been experimentally demonstrated~\cite{gpst_wang_2025}. We are therefore confident that this methodology can be used to construct a diverse range of Hamiltonians tailored for various experimental requirements~\cite{inverse_2017_wang, inverse_2018_chertkov, inverse_2023_inui, inverse_2024_inui}. 
    Nevertheless, there remains room for further improvement. First, the eigenvalue spectrum in our current approach is specified a priori. Understanding how this structure affects the dynamical properties of the Hamiltonian would be of significant interest. Second, the current inverse eigenvalue method imposes strong constraints on the Hamiltonian form (i.e., tridiagonal and mirror symmetric). Investigating whether these constraints can be relaxed to construct more general Hamiltonians is a promising direction for future research.

\begin{acknowledgments}
    This work is supported by the Strategic Priority Research Program of the Chinese Academy of Sciences (Grant No.~XDB0500000), the National Natural Science Foundation of China (Grant No.~12034018 and No.~U23A2074) and the Innovation Program for Quantum Science and Technology (2021ZD0301200, 2021ZD0301500).  
\end{acknowledgments}

\appendix
\section{Inverse eigenvalue method}\label{app:iem}
    We describe how to construct the Hamiltonian parameters $\omega_n$ and $J_n$ that satisfy the PST condition from a given eigenvalue spectrum $\lambda_s (s=1,2,\cdots,N)$. Although efficient algorithms such as the Lanczos algorithm exist for iteratively reconstructing the Hamiltonian from the eigenvalue spectrum and specific elements of the eigenvector matrix (e.g., the first row)~\cite{2006_iep_gladwell, 2017_Lanczos_quantum_Kay, 2017_Lanczos_njp_Kay}, their practical application is limited by the requirement of prior knowledge of the eigenvectors.

    To overcome this limitation, we introduce an inverse eigenvalue method (IEM) based on the theory of orthogonal polynomials, which are closely related to nearest-neighbor (NN) XY-type Hamiltonians (i.e., Jacobi matrices)~\cite{2006_iep_gladwell, 2005_iep_Chu}. To introduce the concept of orthogonal polynomials, we define $\mathbb{P}_N$ as the linear space of all polynomials $p_i(x)$ with degree $i \le N$. An inner product $(\cdot, \cdot)$ is defined on this space, satisfying the properties of positive-definite, bilinear, and symmetric, as follows
    \begin{enumerate}
        \item $(p_i, p_i) \equiv \norm{p_i}^2 > 0$ if $p_i(x) \ne 0$;
        \item $(c p_i, p_j) = c(p_i, p_j)$;
        \item $(p_i + p_j, p_k) = (p_i, p_k) + (p_j, p_k)$; 
        \item $(p_i, p_j) = (p_j, p_i)$;
        \item $(x p_i, p_j) = (p_i, x p_j)$.
    \end{enumerate}

    A special class of orthogonal polynomials, known as monic polynomials, is defined such that the polynomial $p_i(x)$ has degree $i$ and a leading coefficient of one (i.e., the coefficient of $x^i$ is unity). Any two monic polynomials of different degrees in the sequence $\{p_i(x)\}_{i=1}^{N}$ 
    are orthogonal under the inner product  
    \begin{equation}
        (p_i, p_j) = \sum_{s=1}^{N} \alpha_s p_i(x_s) p_j(x_s) = \delta_{ij} \norm{p_i}^2
        . \label{eq-inner}
    \end{equation}
    The weights $\alpha_s$ associated with the inner product are positive and normalized such that 
    \begin{equation}
        \sum_{s=1}^{N} \alpha_s = 1,\quad \alpha_s > 0. \label{eq-weights_norm}
    \end{equation}
    \begin{theorem}
    The most important property of monic polynomials is that they satisfy the three-term recurrence relation, namely 
    \begin{equation}
        p_i(x) = (x - a_i) p_{i-1}(x) - b_{i-1}^2 p_{i-2}(x), \label{eq-monic}
    \end{equation}
    where $i = 1, 2, \cdots, N$. The initial values are given by 
    \begin{equation}
        p_{-1}(x) = 0,\quad p_0(x) = 1. \label{eq-initial}
    \end{equation}
    \end{theorem}
    \begin{proof}
    To prove Eq.~\eqref{eq-monic}, we first move $x p_{i-1}(x)$ to the left-hand side, so that the expression $p_i(x) - x p_{i-1}(x)$ becomes a polynomial of degree $(i-1)$. This term can then be expressed as a linear combination of $p_0, p_1, \cdots, p_{i-1}$ 
    \begin{equation}
        p_i - x p_{i-1} = \sum_{k=1}^{i-1} c_k p_k. \label{eq-pi_comb}
    \end{equation}
    Taking the inner product of both sides with $p_j$ gives
    \begin{equation}
        (p_i, p_j) - (p_{i-1}, x p_j) = \sum_{k=1}^{i-1} c_k (p_k, p_j) = c_j \norm{p_j}^2,
    \end{equation}
    where the left-hand side is simplified using the properties~3 and 5. For $j=1,2,\cdots,i-3$, $x p_j$ has degree at most $i-2$, so the left-hand side vanishes by the orthogonality condition in Eq.~\eqref{eq-inner}. Therefore, $c_j=0$ for all $j$ except $j=i-2$ and $j=i-1$. Eq.~\eqref{eq-pi_comb} reduces to 
    \begin{equation}
        p_i - x p_{i-1} = c_{i-2} p_{i-2} + c_{i-1} p_{i-1}.
    \end{equation}
    To determine $c_{i-1}$ and $c_{i-2}$ (and hence $a_i$ and $b_{i-1}$ in Eq.~\eqref{eq-monic}), we take the inner product of the above equation with $p_{i-1}$, which gives
    \begin{equation}
        a_i = - c_{i-1} = \frac{(p_{i-1}, x p_{i-1})}{\norm{p_{i-1}}^2}, \label{eq-ai}
    \end{equation}
    similarly, taking the inner product with $p_{i-2}$ yields 
    \begin{align}
        c_{i-2} = - b_{i-1}^2 &= -\frac{(p_{i-1}, x p_{i-2})}{\norm{p_{i-2}}^2} \nonumber\\
        &= -\frac{(p_{i-1}, p_{i-1} + \sum_{k=0}^{i-2} c_k' p_k)}{\norm{p_{i-2}}^2} \nonumber\\
        &= -\frac{\norm{p_{i-1}}^2}{\norm{p_{i-2}}^2},
    \end{align}
    where we use the fact that $x p_{i-2}$ is a monic polynomial of degree $(i-1)$ and can therefore be expressed as a linear combination of $p_k$ for $k=0, 1, \cdots, i-1$. Hence, the coefficient $b_{i}$ can be expressed as 
    \begin{equation}
        b_{i} = \frac{\norm{p_{i}}}{\norm{p_{i-1}}}. \label{eq-bi}
    \end{equation}
    \end{proof}
    The three-term recurrence relation is crucial because, together with Eqs.~\eqref{eq-monic}, \eqref{eq-ai} and \eqref{eq-bi}, it enables the full reconstruction of the orthogonal polynomials $\mathbb{P}_N$ starting from the initial conditions Eq.~\eqref{eq-initial}. 

    We now turn our attention to the structure of the NN XY-type Hamiltonian $H$, whose matrix is given in the following tridiagonal form
    \begin{equation}
        H = 
        \begin{pmatrix}
         \omega_1 & J_1 &  &  &  &  & \\
         J_1 & \omega_2 & J_2 &  &  &  & \\
          & J_2 & \omega_3 & J_3 &  &  &\\
          &  &  J_3 & \omega_4 & J_4 &  &\\
          &  &  &  \ddots & \ddots  & \ddots & \\
          &  &  &  & J_{N-2} & \omega_{N-1} & J_{N-1}\\
          &  &  &  &  & J_{N-1} & \omega_N
        \end{pmatrix}.
    \end{equation}
    Instead of focusing on the eigenfunctions, we first consider the principal minors of the matrix $A_N = \lambda I - H$. The principal minor of degree $n$ admits the following Laplace expansion  
    \begin{align}
        P_n(\lambda) = \mathrm{det}(A_n) = \sum_{j=1}^{n} (-1)^{i+j} a_{ij} \mathrm{det}(M_{ij}), \label{eq-Laplace}
    \end{align}
    where $n = 1, 2, \cdots, N$, and the summation over $j$ is equivalent to that over $i$. $A_n$ denotes the submatrix of $A_N$ obtained by retaining the first $n$ rows and columns. $a_{ij}$ is the element in the $i$-th row and $j$-th column of $A_n$. $M_{ij}$ is the submatrix obtained by removing the $i$-th row and $j$-th column from $A_n$. The initial values of $P_n$ for degree $-1$ and $0$ are given by
    \begin{align}
        P_{-1} = 0,\quad P_0 = 1,
    \end{align}
    and for a principal minor of degree $n$, we can explicitly write it as
    \begin{align*}
        P_n(\lambda) &= 
        \begin{vmatrix}
         \lambda - \omega_1 & -J_1 &  &  &  \\
         -J_1 & \lambda - \omega_2 & -J_2 &  &  \\
          &  \ddots & \ddots  & \ddots & \\
          &  & -J_{n-2} & \lambda - \omega_{n-1} & -J_{n-1}\\
          &  &  & -J_{n-1} & \lambda - \omega_n
        \end{vmatrix},
    \end{align*}
    by expanding $P_n(\lambda)$ first along the $n$-th row and then the $(n-1)$-th row or column using Eq.~\eqref{eq-Laplace}, we obtain the following recurrence relation
    \begin{align}
        P_n(\lambda) = (\lambda - \omega_n) P_{n-1}(\lambda) - J_{n-1}^2 P_{n-2}(\lambda). \label{eq-principal}
    \end{align}
    By making the replacement 
    \begin{alignat}{2}
        P_n &\to p_i,\quad &\lambda &\to x, \nonumber\\
        \omega_n &\to a_i, &\quad J_{n-1} &\to b_{i-1}.
    \end{alignat}
    it can be shown that the principal minors of the NN XY-type Hamiltonian exhibit exactly the same structure as the three-term recurrence relation for monic polynomials in Eq.~\eqref{eq-monic}. 
    Favard's theorem states that a sequence of monic polynomials satisfying such a recurrence relation with real $\omega_n$ and positive $J_{n-1}^2$ forms an orthogonal polynomial sequence~\cite{iep_favard_1935,iep_chihara_1978}. Therefore, the principal minors $P_0, P_1, \dots, P_{N-1}$ form a finite monic orthogonal polynomial sequence, while $P_N(\lambda)=\prod_{s=1}^{N}(\lambda-\lambda_s)$ is its terminal polynomial and vanishes at every spectral point $\lambda_s$.

    After establishing the connection between principal minors of $H$ and orthogonal polynomials, the Hamiltonian parameters $\omega_n$ and $J_n$ can be explicitly derived by analogy with Eq.~\eqref{eq-ai} and Eq.~\eqref{eq-bi} 
    \begin{equation}
        \omega_n = \frac{(P_{n-1}, \lambda P_{n-1})}{\left\| P_{n-1} \right\|^2}, \label{eq-recurr_wr}
    \end{equation}
    \begin{equation}
        J_n = \frac{\left\| P_n \right\|}{\left\| P_{n-1} \right\|}. \label{eq-recurr_Jr}
    \end{equation}
    For mirror symmetric Hamiltonians, the weights are given by~\cite{2016_FST_Genest, 2006_iep_gladwell} 
    \begin{align}
        \alpha_s = \frac{d^2}{P_{N-1}(\lambda_s) P_N'(\lambda_s)} = \frac{(-1)^{N+s}d}{P_N'(\lambda_s)}, \label{eq-recurr_weight}
    \end{align}
    where $d$ is a normalization constant used to satisfy the condition in Eq.~\eqref{eq-weights_norm}, and it is independent of the construction of the Hamiltonian parameters. Here, $P_N(\lambda)$ is the characteristic polynomial of $H$, given by $P_N(\lambda) = \prod_{s=1}^N(\lambda - \lambda_s)$, and $P_N'(\lambda_s)$ denotes its derivative evaluated at $\lambda_s$. 
    
    Finally, we establish the relationship between the principal minors and the eigenvectors of the Hamiltonian $H$, whose eigenvalue equation is given by  
    \begin{equation}
        H \ket{\lambda_s} = \lambda_s \ket{\lambda_s}, \label{eq-eigeneq}
    \end{equation}
    the eigenvector $\ket{\lambda_s}$ can be expanded in the basis $\ket{n}$ using Eq.~\eqref{eq-ketn}
    \begin{equation}
        \ket{\lambda_s} = \sum_{n=1}^{N} W_{sn} \ket{n} = \sum_{n=1}^{N} \sqrt{\alpha_s} \chi_n(\lambda_s) \ket{n}. \label{eq-ketlam}
    \end{equation}
    Substituting Eq.~\eqref{eq-ketlam} into Eq.~\eqref{eq-eigeneq} and collecting the coefficients of $\ket{n}$ gives 
    \begin{equation}
       J_n \chi_{n+1}(\lambda_s) = (\lambda_s - \omega_n) \chi_n(\lambda_s) - J_{n-1} \chi_{n-1}(\lambda_s), \label{eq-eigenvec}
    \end{equation}
    where $n=1,2\cdots,N-1$. 
    Starting from the initial conditions $\chi_0 = 0$ and $\chi_1 = 1$, the recurrence relation uniquely defines each polynomial $\chi_n(\lambda)$ for an arbitrary spectral variable $\lambda$.
    Dividing both sides of Eq.~\eqref{eq-principal} by $J_1 J_2 \cdots J_{n-1}$ yields
    \begin{align}
        &J_n\frac{P_n(\lambda)}{J_1J_2\cdots J_n} \nonumber\\
        &=
        (\lambda-\omega_n)
        \frac{P_{n-1}(\lambda)}{J_1J_2\cdots J_{n-1}}
        -
        J_{n-1}
        \frac{P_{n-2}(\lambda)}{J_1J_2\cdots J_{n-2}},
    \end{align}
    comparing the above equation with Eq.~\eqref{eq-eigenvec}, one obtains
    \begin{equation}
        \chi_n(\lambda)
        =
        \frac{P_{n-1}(\lambda)}
        {J_1J_2\cdots J_{n-1}},
        \qquad n=1,2,\ldots,N. 
        \label{eq-vec_prin}
    \end{equation}
    The empty product for $n=1$ is defined as unity. By introducing a positive normalization constant $J_N$, Eq.~\eqref{eq-vec_prin} extends naturally to $\chi_{N+1}(\lambda) = P_N(\lambda)/\prod_{n=1}^N J_n = \prod_{s=1}^N (\lambda-\lambda_s)/\prod_{n=1}^N J_n$. Here $\chi_{N+1}(\lambda)$ serve as a terminal polynomial rather than an additional eigenvector component, and satisfies $\chi_{N+1}(\lambda_s) = 0$. 
    Together with $\alpha_s$ from Eq.~\eqref{eq-recurr_weight}, the coefficients of eigenvectors $W_{sn}=\sqrt{\alpha_s}\,\chi_n(\lambda_s)$ can be fully determined.

    We now briefly outline the procedure for solving the inverse problem. Before the iteration begins, we compute the weights $\alpha_s$ at each spectral point $\lambda_s$ using Eq.~\eqref{eq-recurr_weight}, which will be used for subsequent inner product calculations. Starting from $n=1$, we first compute $\omega_1$ using Eq.~\eqref{eq-recurr_wr}, which depends only on $P_0$ with $P_0=1$. Next, using the known values of $\omega_1, P_0, P_{-1}$, and $J_0$ (with $J_0=0$ as the initial condition), we construct $P_1$ using Eq.~\eqref{eq-principal}. Finally, we use $P_0$ and $P_1$ to determine $J_1$ via Eq.~\eqref{eq-recurr_Jr}. After completing each iteration, we store $\omega_1, P_1$, and $J_1$ for use in the next iteration. The eigenfunctions can be constructed from Eqs.~\eqref{eq-ketlam} and \eqref{eq-vec_prin}. Owing to the mirror symmetry of the target Hamiltonian, only $\lceil N/2 \rceil$ (the ceiling of $N/2$) iterations are needed to compute all undetermined Hamiltonian parameters. 

\section{Derivation of the hidden SWAP dynamics}\label{app:swap}
The behavior of the dome configuration in the limiting cases of $m$ warrants detailed analysis. When $m=0$, the dome configuration naturally reduces to the conventional line model, as shown in Figs.~\ref{fig-fig2}(a) and \ref{fig-fig2}(c), where the coupling strengths follow Eq.~\eqref{eq-Jn} and all qubit frequencies are resonant. 
    Although the large-$m$ limit may not be realizable in experiments, it can reveal important structures hidden in our model. To illustrate this, we follow the method described in Ref.~\onlinecite{hidden_hu_2024}, starting with the case $N =3$. The eigenvalue equation for the second site is given by 
    \begin{equation}
    J_1 \mathfrak{c}_1 + \omega_2 \mathfrak{c}_2 + J_2 \mathfrak{c}_3 = E \mathfrak{c}_2,
    \end{equation}
    where $\mathfrak{c}_i (i=1,2,3)$ denotes the expansion coefficients of the eigenfunction in the Hilbert subspace spanned by sites 1, 2, and 3, and $E$ denotes the corresponding eigenenergy. Since $E$ acts only as a constant energy offset in the derivation of the effective Hamiltonian, we do not distinguish between eigenenergies of different subspaces in the following discussion and denote them uniformly by $E$. Solving the above equation gives 
    \begin{equation}
    \mathfrak{c}_2 = \frac{J_1 \mathfrak{c}_1 + J_2\mathfrak{c}_3}{E - \omega_2}, \label{eq-c2,N=3}
    \end{equation}
    substituting Eq.~\eqref{eq-c2,N=3} into the eigenvalue equations for sites 1 and 3 
    \begin{align}
    \omega_1 \mathfrak{c}_1 + J_1 \mathfrak{c}_2 &= E \mathfrak{c}_1, \\
    J_2 \mathfrak{c}_2 + \omega_3 \mathfrak{c}_3 &= E \mathfrak{c}_3,
    \end{align}
    yields an effective Hamiltonian involving only sites 1 and 3
    \begin{equation}
        \begin{pmatrix}
         \omega_1^\text{eff} & J^\text{eff} \\
         J^\text{eff} & \omega_3^\text{eff}
        \end{pmatrix} 
        \begin{pmatrix}
         \mathfrak{c}_1 \\ \mathfrak{c}_3
        \end{pmatrix} = E 
        \begin{pmatrix}
         \mathfrak{c}_1 \\ \mathfrak{c}_3
        \end{pmatrix},
    \end{equation} 
    where 
    \begin{equation}
    \omega_{1,3}^\text{eff} = \omega_{1} + \frac{J_1^2}{E - \omega_2}, \quad 
    J^\text{eff} = \frac{J_1 J_2}{E- \omega_2}.
    \end{equation}
    By substituting the expressions of $\omega_n$ and $J_n$ from Eqs.~\eqref{eq-dome_wn} and \eqref{eq-dome_Jn} into the above equation and taking the large-$m$ limit, we obtain 
    \begin{align}
    \lim_{m\rightarrow \infty}\omega_{1,3}^\text{eff} &= \omega_{1} + \lim_{m\rightarrow \infty}\frac{J^2(m+1)}{2(E - mJ)} = \omega_{1} - \frac{J}{2}, \nonumber\\
    \lim_{m\rightarrow \infty}J^\text{eff} &= \lim_{m\rightarrow \infty}\frac{J^2(m+1)}{2(E - mJ)} = -\frac{J}{2}.
    \end{align} 
    
    For $N = 4$, using the same approach, we begin with the eigenvalue equations for the central sites 2 and 3
    \begin{align}
        J_1 \mathfrak{c}_1 + \omega_2 \mathfrak{c}_2 + J_2 \mathfrak{c}_3 &= E \mathfrak{c}_2, \\
        J_2 \mathfrak{c}_2 + \omega_3 \mathfrak{c}_3 + J_3 \mathfrak{c}_4 &= E \mathfrak{c}_3,
    \end{align}
    which yields expressions for $\mathfrak{c}_2$ and $\mathfrak{c}_3$ 
    \begin{align}
        \mathfrak{c}_2 &= \frac{J_1(E - \omega_3)\mathfrak{c}_1 + J_2 J_3 \mathfrak{c}_4}{(E - \omega_2)(E - \omega_3) - J_2^2},\\ 
        \mathfrak{c}_3 &= \frac{J_1 J_2 \mathfrak{c}_1 + J_3 (E - \omega_2) \mathfrak{c}_4}{(E - \omega_2)(E - \omega_3) - J_2^2},
    \end{align}
    with these results, we then consider the equations for sites 1 and 4
    \begin{align}
        \omega_1 \mathfrak{c}_1 + J_1 \mathfrak{c}_2 &= E \mathfrak{c}_1, \\
        J_3 \mathfrak{c}_3 + \omega_4 \mathfrak{c}_4 &= E \mathfrak{c}_4,
    \end{align}
    expressing $\mathfrak{c}_2$ and $\mathfrak{c}_3$ in terms of $\mathfrak{c}_1$ and $\mathfrak{c}_4$ leads to a reduced model  
    \begin{equation}
        \begin{pmatrix}
         \omega_1^\text{eff} & J^\text{eff} \\
         J^\text{eff} & \omega_4^\text{eff}
        \end{pmatrix} 
        \begin{pmatrix}
         \mathfrak{c}_1 \\ \mathfrak{c}_4
        \end{pmatrix} = E 
        \begin{pmatrix}
         \mathfrak{c}_1 \\ \mathfrak{c}_4
        \end{pmatrix},
    \end{equation}
    with 
    \begin{align}
    \omega_n^\text{eff} &= \omega_n + \frac{J_1^2 (E - \omega_3)}{(E-\omega_2)(E-\omega_3) - J_2^2}, \quad n =1, 4,\\
    J^\text{eff} &= \frac{J_1 J_2 J_3}{(E - \omega_2)(E - \omega_3) - J_2^2},
    \end{align}
    in the large-$m$ limit, we have
    \begin{align}
        \lim_{m\rightarrow \infty}\omega_{1,4}^\text{eff} = \omega_{1} - J, \quad 
        \lim_{m\rightarrow \infty}J^\text{eff} = \frac{J}{2}.
    \end{align}
    
    Applying the same approach to $N=5$, we iteratively eliminate the central sites from $\mathfrak{c}_2$ to $\mathfrak{c}_4$, expressing them in terms of $\mathfrak{c}_1$ and $\mathfrak{c}_5$, and obtain an effective Hamiltonian involving only sites 1 and 5 
    \begin{equation}
        \begin{pmatrix}
         \omega_1^\text{eff} & J^\text{eff} \\
         J^\text{eff} & \omega_5^\text{eff}
        \end{pmatrix} 
        \begin{pmatrix}
         \mathfrak{c}_1 \\ \mathfrak{c}_5
        \end{pmatrix} = E 
        \begin{pmatrix}
         \mathfrak{c}_1 \\ \mathfrak{c}_5
        \end{pmatrix},
    \end{equation}
    with
    \begin{align}
        \omega_n^\text{eff} &= \omega_n + \frac{(E - \omega_3) J_1^2 \mathcal{A}_2}{\mathcal{A}_1 \mathcal{A}_2 -(J_2J_3)^2}, \quad n=1, 5\\
        J^\text{eff} &= \frac{(E - \omega_3) J_1 J_2 J_3 J_4}{
       \mathcal{A}_1 \mathcal{A}_2 -(J_2J_3)^2},
   \end{align}
   where $\mathcal{A}_1 = (E - \omega_2)(E - \omega_3) - J_2^2$ and $\mathcal{A}_2 = (E - \omega_3)(E - \omega_4) - J_3^2$.  In the large-$m$ limit, we have $\mathcal{A}_1 = \omega_2\omega_3 - J_2^2$, and $\mathcal{A}_2 = \omega_3 \omega_4 - J_3^2$. Using the expressions of $\omega_n$ and $J_n$ and taking the large-$m$ limit, we obtain
   \begin{align}
        \lim_{m\rightarrow \infty}\omega_{1,5}^\text{eff} = \omega_{1} - \frac{3}{2}J, \quad 
        \lim_{m\rightarrow \infty}J^\text{eff} = - \frac{J}{2}.
    \end{align}
    The numerical results for $\omega_{1, N}^\text{eff}$ and $J^\text{eff}$ based on perturbative Schrieffer--Wolff transformation are presented in Fig.~\ref{fig-fig3}~\cite{swt_magesan_2020}. We find that  
    \begin{align}   
        \lim_{m\rightarrow \infty}\omega_{1,N}^\text{eff} &= \omega_{1} - \frac{N-2}{2}J, \\
        \lim_{m\rightarrow \infty}J^\text{eff} &= (-1)^{N} \frac{J}{2} \label{eq-JefflargeN},
    \end{align}
    therefore, by integrating out the intermediate sites $n = 2, 3, \cdots, (N-1)$, we obtain the following effective Hamiltonian in the $m \rightarrow \infty$ limit 
    \begin{equation}
        H^{\text{eff}} = \begin{pmatrix}
            \omega_1^\text{eff} & J^{\text{eff}} \\
            J^{\text{eff}} & \omega_N^\text{eff}
        \end{pmatrix}. \label{eq-Heff}
    \end{equation}
In this limit, PST reduces to a SWAP model between the left- and right-end sites~\cite{2019_coupler_krantz}. This is a direct generalization of the approach in Ref.~\onlinecite{hidden_hu_2024} to the effective coupling between distant sites, where in the large-$m$ limit, the energy $E$ becomes irrelevant. Here, we use this approach to unveil the hidden mathematical structure of our model. 

    In summary, the dome model converges to two distinct effective models in different asymptotic regimes. For $m=0$, the $N$-qubit dome model becomes equivalent to the $N$-qubit line model. As $m \rightarrow \infty$, it reduces to a two-qubit SWAP model between the edge qubits.

\normalem
\bibliography{ref.bib}
\clearpage

\end{document}